\documentclass{aa}  
\usepackage{graphicx}
\usepackage{txfonts}
\usepackage[dvipsnames]{xcolor}
\usepackage{hyperref}
\hypersetup{
    colorlinks=true,
    linkcolor=olive,
    citecolor=RoyalBlue,
    filecolor=magenta,      
    urlcolor=RoyalBlue,
    pdfpagemode=FullScreen,
    }

\newcommand{\rev}[1]{\textcolor{black}{#1}}
\newcommand{\orcit}[1]{\protect\href{https://orcid.org/#1}{\protect\includegraphics[width=8pt]{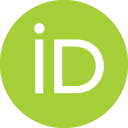}}}
\usepackage[dvipsnames]{xcolor}

\graphicspath{{./}{figures/}}

\begin{document}

\title{Confronting sparse Gaia DR3 photometry with TESS for a sample of around 60,000 \rev{OBAF-type} pulsators}

\author{Daniel Hey\inst{1\,\orcit{0000-0003-3244-5357}} \and
    Conny Aerts\inst{2,3,4\,\orcit{0000-0003-1822-7126}}}

\institute{
Institute for Astronomy, University of Hawai'i, Honolulu, HI 96822, USA\\
email:\ {dhey@hawaii.edu}
\and
Institute of Astronomy, KU\,Leuven, Celestijnenlaan 200D,
B-3001  Leuven, Belgium\\
email:\ {Conny.Aerts@kuleuven.be}
\and
Department of Astrophysics, IMAPP, Radboud University Nijmegen, PO Box 9010,
6500 GL, Nijmegen, The Netherlands
\and
Max Planck Institute for Astronomy, Koenigstuhl 17, 69117, Heidelberg, Germany}

\date{Received February ??, 2024; Accepted ??, 2024}

\abstract
{The Gaia mission has delivered hundreds of thousands of variable star light curves in multiple wavelengths. Recent work demonstrates that these light curves can be used to identify
    (non-)radial pulsations
    in the OBAF-type stars, despite the irregular cadence and low light curve precision
    of order a few mmag. With the considerably more precise TESS photometry, we revisit these candidate pulsators to conclusively ascertain the nature of their variability.}
{We seek to re-classify the Gaia light curves with the first two years of TESS photometry for a sample of 58,970 p- and g-mode pulsators, encompassing $\gamma$ Dor, $\delta$ Scuti,
    SPB, and $\beta$ Cep variables. From the TESS data, we seek to assess the quality of Gaia's classification of non-radial pulsators which is based on sparse, years-long light curves of mmag precision. We also supply four new catalogues containing the confirmed pulsators, along with their dominant and secondary pulsation frequencies, the number of independent mode frequencies, and a ranking according to their usefulness for future asteroseismic ensemble analysis.}
{We first analyze the TESS light curves independent of their Gaia classification by prewhitening all dominant pulsation modes down to a 1\% false alarm probability. Using this, in combination with a feature-based random forest classifier, we identify different variability types across the sample.}
{We find that the Gaia photometry is exceptionally accurate for detecting the dominant and secondary frequencies, reaching approximately 80\% accuracy in frequency for p- and g-mode pulsators.
    The majority of Gaia classifications are consistent with the classifications from the TESS data,
    illustrating the power of
    the low-cadence Gaia photometry for pulsation studies.
    We find that the sample of g-mode pulsators forms a continuous group of variable stars
    along the main sequence across
    B, A, and F spectral types, implying that
    the mode excitation mechanisms for all these pulsators need to be updated with improved physics.
    Finally, we provide a rank-ordered table of pulsators according to their asteroseismic potential for follow-up studies, based on the number of sectors they have been observed in, their classification probability,
    and the number of independent modes found in the TESS light curves from the nominal mission. }
{Our catalogue offers a 
major increase in the number of confirmed gravity-mode pulsators with an identified dominant mode suitable for follow-up TESS ensemble asteroseismology of such stars.}


\titlerunning{confronting Gaia DR3 with TESS light curves of OBAF-type nonradial pulsators}
\authorrunning{Hey \& Aerts}

\maketitle
%

\section{Introduction}

Over the course of the past century, variable stars have proven to be an invaluable resource in the refinement of our understanding of stellar structure and evolution. This has become notably apparent following the advent of high-precision space photometry, a revolution which has provided astronomers with a continuous stream of high-frequency micro-magnitude precision light curves spanning the entirety of the Hertzsprung-Russell diagram \citep[HRD,][]{Kurtz2022}. The richness of this data has paved the way for advanced asteroseismic investigations into the internal structure of stars \citep{Aerts2021-RMP}.

Asteroseismic modelling has been utilized for thousands of red giants, with stochastically excited identified modes exhibiting amplitudes around 0.1\,mmag and periodicities on the order of hours \citep[e.g.,][for a review]{HekkerJCD2017}. Conversely, the sample of uniformly analysed white dwarf successors to these red giants only includes a few tens of compact pulsators with identified modes \citep[e.g.][]{Hermes2017}. Despite the possibility of their amplitudes reaching several mmag, \rev{the limited sample size of DAV and DBV white dwarf pulsators is primarily due to their characteristics as faint, fast gravity-mode pulsators \citep[e.g.][for a summary]{Corsico2019}.} Therefore, monitoring these pulsators for asteroseismology necessitates a cadence under a minute, as opposed to the half-hour cadences suitable for covering the hours-long periodicities of red giant modes.

Asteroseismology of main sequence pulsators has so far also only been applied to limited ensembles compared to red giants. For low-mass dwarfs like the Sun this limitation is due to the low amplitudes (of order 10\,$\mu$mag) and short periods (of order minutes) of their solar-like pulsations \citep[][for a review]{GarciaBallot2019}. For intermediate- and high-mass dwarfs the limitations are mainly caused by the fast rotation \citep[see][for a review]{AertsTkachenko2024}, preventing identification of the asymmetrically split merged mode multiplets for the majority of discovered class members. So far, homogeneous asteroseismic modelling of intermediate-mass dwarfs, treating all pulsators in the same way has only been done for a few modest ensembles:
\begin{itemize}
    \item
          60 young unevolved $\delta\,$Scuti stars very close to the zero-age main sequence with high-frequency pressure modes \citep{Bedding2020}. As for most of the slowly rotating red giant pulsators, their modelling was done while ignoring the Coriolis force and relied on just the large frequency separation  and the frequency of maximum power \citep{Panda2024}, rather than the fitting of individual mode frequencies;
    \item
          26 slowly pulsating B stars \citep[SPB stars,][]{Pedersen2021} whose individual identified mode frequencies were modelled based on the methodological framework designed specifically for gravito-inertial modes in fast rotators relying on the Traditional Approximation of Rotation (TAR), as developed in \citet[][to which we refer for details]{Aerts2018};
    \item
          490 $\gamma\,$Doradus ($\gamma\,$Dor) stars whose measured buoyancy travel time and near-core rotation frequency deduced from identified gravito-inertial modes adopting the TAR were modelled by \citet{Mombarg2024};  37 of these $\gamma\,$Dor stars had their individual modes modelled by methodology based on the TAR coupled to a neural network \citep{Mombarg2021}.
\end{itemize}
For the high-mass $\beta\,$Cep stars, homogeneous ensemble modelling of their few (typically between 3 and 5) identified pressure modes has not yet been done; \citet{Bowman2020-FrASS} summarised some of the results for seven individual class members. Clearly, ensemble asteroseismology of intermediate- and high-mass dwarfs is still in its early stages, despite the discovery of thousands of class members \citep{Handler2019Asteroseismology, Burssens2023Calibration, Eze2024Beta}. Lack of mode identification prevents applications to large ensembles. Novel approaches to filter thousands of main-sequence pulsators with proper mode identification are therefore in order. The initial attempts for the toughest case of gravito-inertial pulsators by \citet{Garcia2022b,Garcia2022a} based on just the first year of TESS monitoring already illustrated the major potential of this mission for ensemble asteroseismology of fast rotators across the Milky Way, in line with the opportunities discussed in \citet{AertsTkachenko2024}.

The current study aims to increase the sample of main sequence pulsators of spectral types O, B, A, and F (hereafter OBAF pulsators) with the potential for ensemble asteroseismology with an order of magnitude. Our work is focused on the availability of homogeneously assembled seismic and non-seismic observables. Even if it was not designed for this type of research, the ESA Gaia mission \citep{Prusti2016,Brown2016} has a significant role to play in this context. We tackle the challenging search for suitable large ensembles of dwarf pulsators by screening the homogeneous database of stellar properties and sparse time-series photometry offered by Gaia's Data Release\,3 \citep[DR3,][]{Vallenari2023}.
Starting from the sample of 106,207 candidate OBAF pulsators classified by \citet{DeRidder2023},
we analyse those targets among this large ensemble having high-cadence high-precision light curves assembled by the NASA TESS space telescope \citep{Ricker2015}.

In order to prepare the `industrialisation' of asteroseismic ensemble modelling of OBAF pulsators, we need to find targets with tens of identified modes from the TESS photometry among the `zoo of variables' along the main sequence \citep{Eyer2023}. Thousands of variable intermediate- and high-mass dwarfs have already been found from high-cadence uninterrupted space photometry
\citep[e.g.,][]{Uytterhoeven2011,Balona2011-Ap,Balona2011-DSCT,Balona2011-Btype,Balona2016,Balona2019,Antoci2019,Pedersen2019,Burssens2019,Burssens2020}. Their variability is caused by different physical phenomena, making identification of the pulsation modes for large ensembles of OBAF pulsators a major obstacle. In this era of high-cadence space photometry the variability phenomena are better understood and include stellar flaring \citep[e.g.][]{Balona2015,Pedersen2017} along with other magnetic variability \citep[e.g.][]{Shultz2019,David-Uraz2019,Sikora2019a}, rotational modulation \citep[e.g.,][]{Bowman2018,Sikora2019b}, low-frequency variability interpreted in terms of internal gravity waves \citep[e.g.,][]{Aerts2015,Bowman2019,Bowman2020} or subsurface convection \citep[e.g.,][]{Cantiello2019,Cantiello2021}, eclipses in close binaries \citep[e.g.,][]{Kirk2016,IJspeert2021} and so on. This non-exhaustive list of variable phenomena often occurs in combination with nonradial pulsations \citep[as summarised in][]{Kurtz2022}.

Furthermore, it has long been established that OBAF pulsators coexist with various types of variable stars along the main sequence in the Hertzsprung-Russell diagram (HRD) \citep[e.g.,][]{Briquet2007}. The recently released Gaia Data Release 3 (Gaia DR3) data reaffirm this observation, demonstrating that the classes of variability, including OBAF pulsators, encompass a substantial proportion of stars distributed across nearly the entire main sequence within the intermediate- and high-mass dwarf star population \citep{DeRidder2023}. This phenomenon is further substantiated by investigations of young open clusters, which have been systematically monitored through joint efforts involving the Gaia mission, as well as the refurbished {\it Kepler\/} (known as K2) and Transiting Exoplanet Survey Satellite (TESS) missions \citep[e.g.,][]{White2017,Murphy2022,Bedding2023,Fritzewski2024,GangLi2024}. Consequently, it is prudent to adopt an approach that involves scrutinizing the time-series photometric data itself, rather than solely focusing on the stellar positions within the HRD, to effectively address our scientific objectives. Gaia DR3 provides us with the requisite data to pursue this approach in our quest to identify the most appropriate ensembles of OBAF pulsators.

In this paper, we revisit a large fraction of the 106207 OBAF pulsators discovered in the sparse Gaia DR3 photometry by \citet{DeRidder2023} with the goal to derive and analyse their TESS light curves. We focus on the poorly populated samples of nonradial pulsators having the highest asteroseismic potential and thus revisit the candidate $\gamma\,$Dor stars, Slowly pulsating B (SPB) stars, $\beta\,$Cep stars, and $\delta$ Scuti stars identified by \citet{DeRidder2023}. The $\gamma\,$Dor and SPB g-mode pulsators have been revisited already by \citet{Aerts2023Astrophysical} in terms of their astrophysical properties, but the $\beta\,$Cep and $\delta$ Scuti p-mode pulsators have not been evaluated as such.  Here, we confront the Gaia DR3 variability properties for the members assigned to these four classes with their TESS data when available. Our work has the following two major aims:
\begin{itemize}
    \item
          to assess the quality of Gaia's classification of nonradial pulsators, which is only based on sparse years-long light curves of mmag precision, by confronting its outcome with high-cadence $\mu$mag precision TESS space photometry for all class members that have Gaia and TESS independent data sets in the public domain;
    \item
          to create four large new catalogues containing Gaia-discovered nonradial pulsators confirmed by TESS, offering the community their TESS light curves covering the first year of monitoring, their independent mode frequencies, and identification of the dominant mode if possible. These catalogues are an optimal starting point for future ensemble asteroseismic modelling.
\end{itemize}

\section{Gaia and TESS Data}

\begin{figure*}[t!]
    \begin{center}
        \includegraphics[]{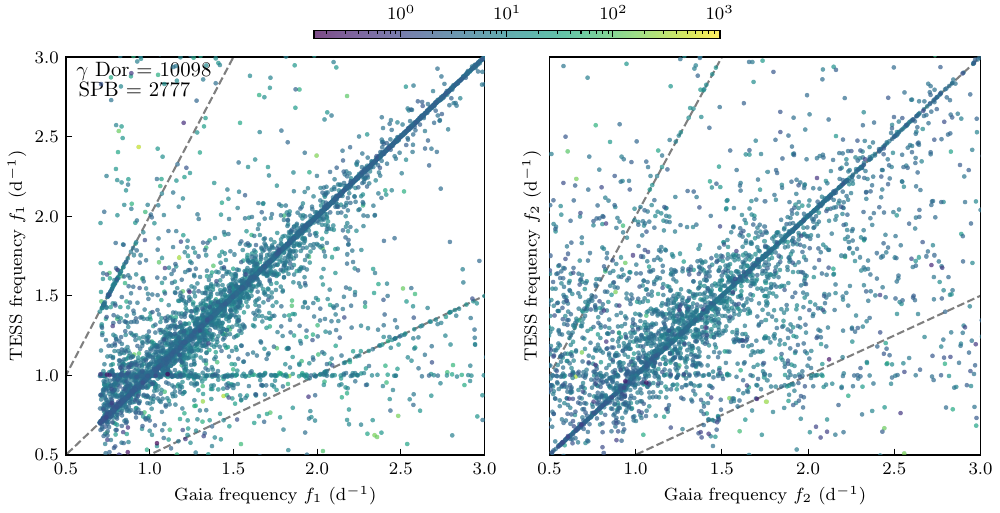}
        \includegraphics[]{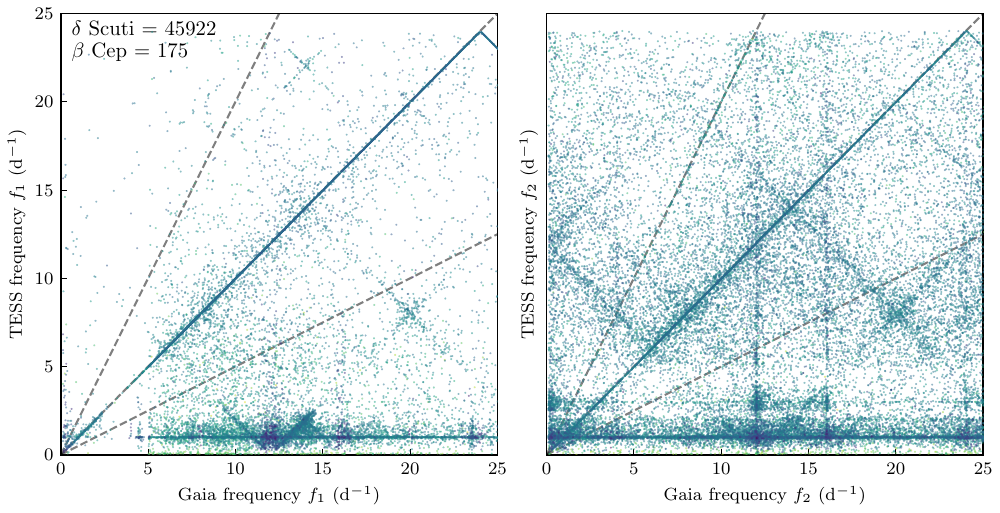}
    \end{center}
    \caption{Left: \rev{Comparison of measured dominant TESS frequency against \textit{Gaia} frequency for the g-mode pulsators (top) and p-mode pulsators (bottom), colored by the amplitude of the variability (ppt). Right: The same for the secondary frequency. The dashed lines indicate the half, unity, and twice relationships among the frequencies. We compare the sample of g-mode candidate pulsators before removing combination frequencies, to match the method of \citet{Aerts2023Astrophysical}. The criss-cross structures in the p-mode plots are caused by aliases of the true signal around the Nyquist limit.}}
    \label{fig:freqfreq}
\end{figure*}

\begin{figure*}[t]
    \begin{center}
        \includegraphics[scale=1]{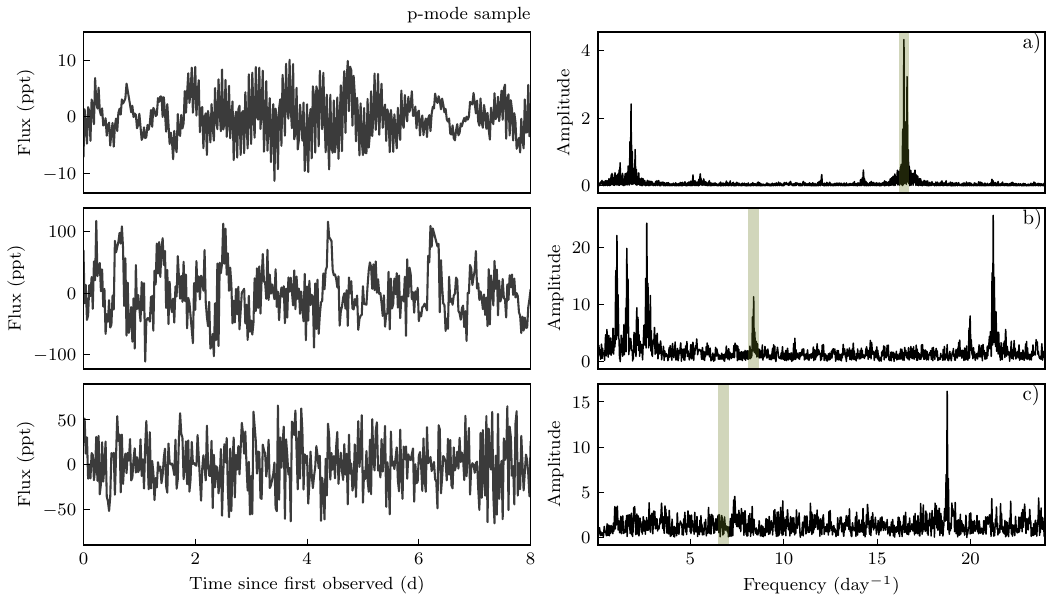}
        \includegraphics[scale=1]{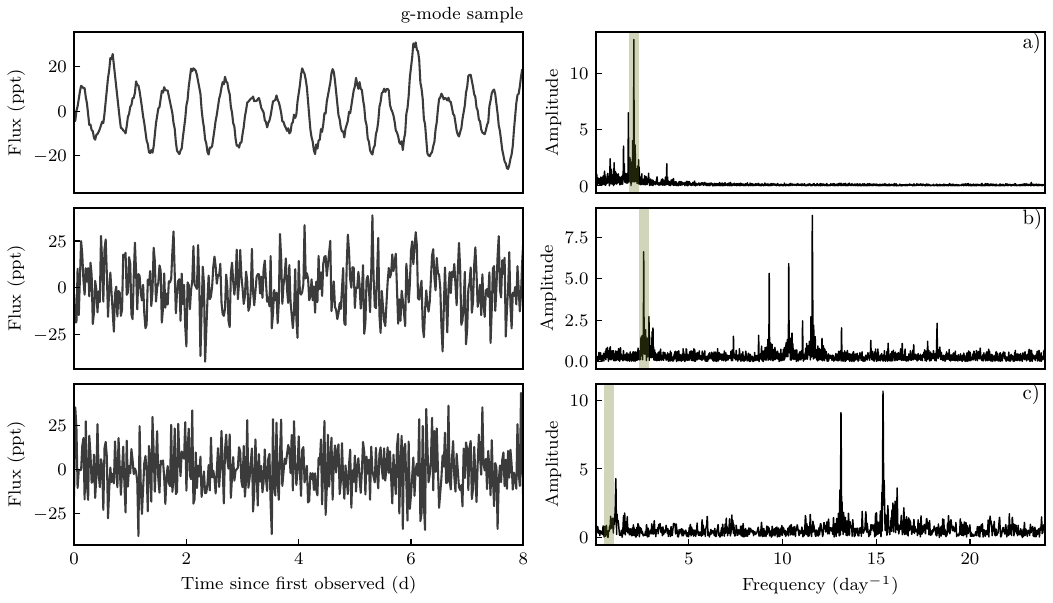}
    \end{center}
    \caption{
        Example light curves and amplitude spectra from the p-mode (top three panels) and g-mode (bottom three panels) candidate sample. The green vertical line marks the measured Gaia frequency. We show the following cases; a) where the measured Gaia frequency is in good agreement with TESS, b) where the measured Gaia frequency is not the true dominant one, and c) the measured Gaia frequency is incorrect.
    }
    \label{fig:LCs}
\end{figure*}

\begin{figure*}[t!]
    \begin{center}
        \includegraphics[]{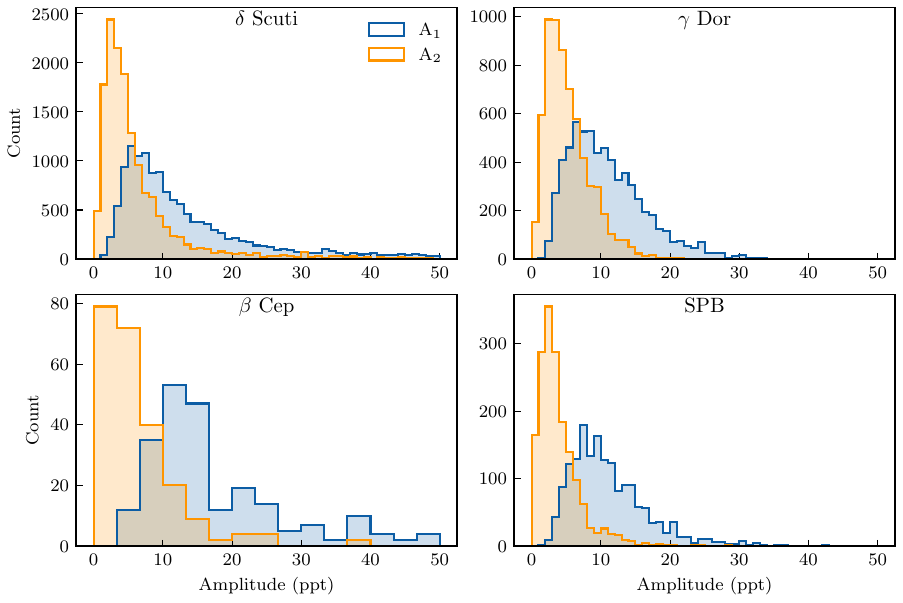}
    \end{center}
    \caption{Distribution of amplitudes for the dominant (blue) and secondary (orange) frequencies for each variable type in our sample. Note that we use only stars where the dominant frequency lies on the bisector within 0.1 d$^{-1}$ tolerance for the TESS/Gaia overlap (cf. Fig.~\ref{fig:freqfreq}).}
    \label{fig:amp_dist}
\end{figure*}

\subsection{The four samples of candidate pulsators}

Our samples consist of the p- and g-mode pulsators discussed in \citet{DeRidder2023}.
For the g-mode pulsators, we take the 15,062 candidates from \citet{Aerts2023Astrophysical}. This sample contains 11,636 $\gamma$\,Dor and 3,462 SPB star candidates. In addition, we consider the 222 candidate $\beta\,$Cep stars and 85,317 $\delta$\,Scuti candidates classified as such by \citet{DeRidder2023}. For the latter two classes, the extra vetting based on the expected frequency range as done for the g-mode pulsators by \citet{Aerts2023Astrophysical} is not meaningful, because their dominant p-mode frequencies are expected to intervene with (multiples of) Gaia's main satellite frequency near 4\,d$^{-1}$ at mmag level. Moreover, large fractions among the $\beta\,$Cep and $\delta\,$Scuti stars may have a dominant high-amplitude radial mode, so a restriction on amplitude as extra vetting rule as adopted by \citet{Aerts2023Astrophysical} for the $\gamma\,$Dor and SPB pulsators is less obvious to define for the p-mode pulsators.

To construct the four samples for the current work, we cross-match their Gaia Data Release 2
\citep[DR2,][]{Brown2018}
identifications (IDs) using the `nearest neighbours' table. To obtain their TESS IDs, we then cross-match the Gaia DR2 IDs against the TESS Input Catalog \citep[TIC,][]{Stassun2018TESS}. The final cross-matched sample among DR3 \citep{Vallenari2023}, DR2, and TIC contains 85,313 $\delta$\,Scuti stars, 11,636 $\gamma$\,Dor stars, 3,426 SPB stars, and 222 $\beta$ Cep stars. The loss of several stars in the cross-matching process is a result of the DR2 to DR3 best neighbours matching catalogue, which is not strictly one-to-one.

\subsection{TESS photometry}

The majority of the sample has not been targeted for dedicated observations by TESS.
With no official light curves delivered by the TESS team using the SPOC pipeline, we instead use light curves from the TESS Gaia light curve (TGLC) catalogue, produced from full-frame images by \citet{Han2023TESSGaia}. TGLC light curves were used as an alternative to the Quick-Look Pipeline (QLP, \citet{Huang2020Photometry} because they reach fainter than 13.5 mag. The TGLC light curves leverage Gaia photometry and astrometry to inform light curve extraction by building a local point spread function forward model, making it a viable source for fainter stars, where Gaia astrometry is useful for resolving contaminated stars.

We use the TGLC light curves from the first 26 sectors of the TESS mission calculated by the TGLC pipeline from the full-frame images. These light curves are at a nominal 30-minute cadence. We use the calibrated aperture flux with the ``good'' quality flags. For each light curve, we perform 3$\sigma$ clipping of the flux and apply a Gaussian filter with a standard deviation of 100 to remove significant long-term trends. A table containing the curl scripts used to download the data is available in electronic format as supplementary material. The TGLC light curves are available in the first 26 TESS sectors for 45,919 $\delta$ Scuti stars, 10,099 $\gamma$ Doradus stars, 2,777 SPB stars, and 175 $\beta$ Cep stars, leading to a total analysis sample of 58,970 variables out of the original target list. 

\section{Confrontation between the two light curves per sample star}

\subsection{Prewhitening of dominant pulsation modes}

To analyze the candidate pulsators in our sample we have developed an automated prewhitening scheme based on the approach used for p-mode pulsators in \citet{Hey2021Searcha}, with several modifications. The algorithm functions as follows: it begins by identifying the highest peak in the amplitude spectrum. It then optimises the value of the peak frequency from a sinusoidal function in the time domain characterized by a frequency, amplitude, and phase. This fitted sinusoid is then subtracted from the light curve, repeating iteratively until a predefined stopping condition.
In contrast to \citet{Hey2021Searcha} where the stopping condition was the signal-to-noise ratio (SNR) of the peak, our method employs the false-alarm level for the highest amplitude peak, ensuring it falls below a 1\% probability threshold. If this condition is met, the peak is deemed significant, removed, and the prewhitening process continues.

The prewhitening procedure is applied to each peak exceeding the 1\% significance level in the amplitude spectrum, or until a maximum of 100 iterations is reached, whichever occurs first. Additionally, for each peak, we calculate its `prominence', which quantifies the peak's distinctiveness relative to the surrounding amplitude spectrum. This metric serves as a useful diagnostic tool for evaluating individual modes, especially in the scenario where a single prewhitening step does not completely remove a peak (a common occurrence in non-sinusoidal signals).

We perform the prewhitening for all stars in our sample. Stars with more than one sector of TESS observations are stitched together prior to prewhitening, with each sector having a simple 5$\sigma$ outlier removal and long-term trend removal with a Gaussian filter. During the prewhitening routine, it is common to encounter combination frequencies \citep{Kurtz2015unifying}. These can be identified and subsequently removed using the following equation \citep{GangLi2019}:
\begin{equation}
    |f_k - (n_i f_i + n_j f_j) | < \epsilon\, ;
\end{equation}
where $i$, $j$, and $k$ are indices of the frequency peaks, $f_i$ and $f_j$ are the parent frequencies, $f_k$ is the candidate for the combination frequency, $n_i$ and $n_j$ are combination coefficients, and $\epsilon$ is the threshold for identification. Following the approach of \citet{GangLi2019}, we limit our analysis to the 20 highest amplitude peaks, considering them as potential parent frequencies. Our criterion for combinations is restricted to cases where $|n_i| + |n_j| \leq 2$. In light of the TESS data's lower precision compared to the {\it Kepler\/} data, we opt for a considerably larger $\epsilon$ value of 0.002~d$^{-1}$, compared to that used by \citet{GangLi2019}. We also remove harmonics, defined as integer or half-integer multiples of the parent frequency. Such harmonics are common, for example, in eclipsing binary samples \citep[e.g.,][]{IJspeert2024}.

\subsection{Comparison of the two dominant frequencies}

Here we compare the results of our TESS light curves against the Gaia photometry explored in
\citet{DeRidder2023} and \citet{Aerts2023Astrophysical}. Figure~\ref{fig:freqfreq} shows the comparison between dominant and secondary modes between the TESS and Gaia data for all the candidate pulsators in the four samples, grouping the p-mode and g-mode pulsators in separate panels. The agreement is particularly good considering the sparse sampling of the Gaia data, with 69\% of the sample lying along the bisector with a 0.1~d$^{-1}$ tolerance for the dominant frequencies of the p-modes, and 80\% for the g-mode frequencies. This result on comparisons between the dominant frequencies in the Gaia and TESS light curves is superior to the 20\% agreement in dominant frequency between the Gaia and {\it Kepler\/} light curves for the few tens of  $\gamma\,$Dor and SPB stars found by \citet[][see their Appendix\,A]{DeRidder2023}.

There is a clear systematic in the TGLC light curves at 1~d$^{-1}$ caused by what we believe to be reflected light from Earth (`earthshine'). We find no correlation between the amplitude of this signal and whether the star falls along a TESS bisector or not.
\rev{For the p-mode pulsators, Fig.\,\ref{fig:freqfreq} reveals additional criss-crossing structures aside from the 1~d$^{-1}$ systematic in the TESS data. This phenomenon is understood in terms of the 30-min sampling by TESS, causing a mirroring effect around the Nyquist frequency that leads to an aliased signal of the true frequency. For example, a $\delta$ Scuti variable with a true pulsation at 44 d$^{-1}$ will have an indistinguishable copy of the signal appearing at around 4 d$^{-1}$ if observed in 30-minute cadence. This effect can be seen, for example, in the $\delta$ Scuti and rapidly oscillating Ap stars observed by {\it Kepler\/} in long-cadence \citep{Bell2017Destroying,Murphy2019Gaiaderived,Hey2019Six}. The clustering of modes around 10-15 d$^{-1}$ are likely aliases of the true $\delta$ Scuti p-mode oscillations that are expected to peak above the Nyquist limit \citet{Hey2021Searcha}. Note that the true and aliased signals of coherent pulsators can be distinguished in the {\it Kepler\/} data as a consequence of periodic modulation of the light curve measurement times \citep{Murphy2013SuperNyquist}.}

We show a series of light curves in Fig.~\ref{fig:LCs} to demonstrate pathological cases of the Gaia and TESS data. For the p- and g-mode candidate sample, we illustrate three occurring scenarios: a) the Gaia measured dominant frequency is correct and confirmed by TESS, b) Gaia picks up a secondary peak of an otherwise correctly classified pulsator, and c) the dominant Gaia frequency is wrong and likely of instrumental origin. The latter situation calls for a re-evaluation of the Gaia DR3 variability classification of the star, based on its TESS light curves. We tackle this subject in Sect.\,4. We also note that the Gaia identification of dominant and secondary frequencies is dependent on the scanning law \citep{Steen2024Measuring}.

Figure~\ref{fig:amp_dist} shows the histograms of the primary and secondary amplitudes for the TESS data of each variable type, where the dominant frequency is in good (that is, to within 0.1~d$^{-1}$) agreement with the measured Gaia frequency, and the classification is accurate in both TESS and Gaia with a classification probability $>0.5$ (Section \ref{sec:classification}).
We perform a two-sided Kolmogorov-Smirnov test to compare the amplitude distributions across each of the two variability classes with the same type of modes. The null hypothesis for this test is that the two distributions are identical. That is, they are drawn from the same underlying distribution. We choose a confidence level of 95\% for the test, such that the null hypothesis is rejected if the p-value is less than 0.05. 

For the $\delta$\,Scuti and $\beta$\,Cep sample, the test indicates that their amplitude distributions are statistically different, with a p-value of around
$10^{-14}$. 
\rev{
This result can be understood from various arguments. First, the two classes of p-mode pulsators are subject to the same type of excitation mechanism, namely the opacity (or $\kappa$) mechanism, but  
it acts in a different layer in the outer envelope of the star. For the $\beta\,$Cep stars it concerns the partial ionisation zone of iron-like isotopes occurring at a temperature of about 200kK, while the heat engine in the case of $\delta\,$Scuti stars is mainly acting in the cooler second ionization zone of helium located in the region with a temperature of about 45kK, with a small contribution from the even cooler hydrogen ionization zone as well
\citep{Pamyatnykh1999, Guzik2021Highlights}. Moreover, in the $\beta\,$Cep stars the excitation layer is situated within a fully radiative envelope, while the $\delta\,$Scuti stars have a thin outer convective envelope where proper treatment of time-dependent convection is required to perform the mode calculations \citep{Dupret2005Convectionpulsation}. These different zones have different local time scales and hence lead to different mode periodicities and amplitudes. The wave damping is also different in the outer layers of both types of pulsators. Finally, unlike for the $\beta\,$Cep stars, strong amplitude modulation occurs in many $\delta\,$Scuti stars \citep{Bowman2016Amplitude}. 
}

For the $\gamma$ Dor and SPB sample, however, we find that the null hypothesis can not be rejected with a p-value of 0.055. This indicates that the distributions of their amplitudes are statistically the same. From the observational side, this is expected from prior work based on variability classification -- only colour information allows us to distinguish between the SPB and $\gamma$ Dor classes \rev{from photometric data} \citep{Audenaert2021TESS}, and even then, there is still confusion for stars with an effective temperature between 8500-9500\,K \citep{Aerts2023Astrophysical}. 
\rev{The mode frequency ranges for both types of pulsators are roughly the same, as are the ranges of the overtones of the excited modes \citep{GangLi2020,Pedersen2021} and of the internal rotation rates when expressed in terms of the critical break-up frequency \citep{VanReeth2016,VanReeth2018,Pedersen2022b}.}
The equal amplitude distributions among $\gamma$\,Dor and SPB stars are a natural consequence of the known mode excitation mechanisms
\rev{for the two classes of pulsators. 
While the SPB stars have their g~modes excited by the $\kappa$ mechanism acting in
the same partial ionisation zone of iron-like isotopes as the modes in the $\beta\,$Cep stars \citep{Pamyatnykh1999,Townsend2005,Szewczuk2017}, the $\gamma\,$Dor stars are subject to at least two not mutually exclusive excitation mechanisms: while the $\kappa-$mechanism comes into play for the hottest class members, the dominant excitation is caused by flux blocking at the base of the thin outer convective envelope. This base is situated in a region with temperatures between roughly 200~kK and 500~kK 
\citep{Guzik2000,Dupret2005Timedependent,Bouabid2013} and explains the similarity of the mode frequencies and amplitudes for SPB and $\gamma\,$Dor stars. }

\rev{With our work, we re-affirm that the current excitation predictions cannot be complete as we observe more modes than predicted and they occur also in stars outside the currently available instability strips. This was found from the Gaia DR3 results in \citet{DeRidder2023} and already independently found and confirmed from {\it Kepler\/} and TESS space photometry \citep[cf.\,the recent assembly and summary by ][]{Balona2024}. 
\citet{Fritzewski2024} and \citet{Mombarg2024} found the majority of the observed {\it Kepler\/} $\gamma\,$Dor stars to have a mass above 1.5\,M$_\odot$ and many of them to be hotter than the blue edge of current $\gamma\,$Dor instability strips, well into the $\delta\,$Scuti classical instability strip and even above it.  One promising route to achieve extra mode excitation in addition to the known mechanisms active in SPB and $\gamma\,$Dor stars is
nonlinear mode coupling, which was already established in large-amplitude SPB stars \citep{VanBeeck2021}. This mechanism} may excite extra modes via energy exchange between modes, aside
from the self-driven linear modes due to the $\kappa$ mechanism. While a similar study on nonlinear mode coupling has not yet been done for $\gamma\,$Dor stars, their {\it Kepler\/} light curves show similar cusp-like shapes near the maxima than the SPB stars with nonlinear mode coupling do \citep[cf.\,compare the data in ][]{VanReeth2015,Pedersen2021}.

Finally, it has been shown that adding novel physical ingredients in mode excitation computations may appreciably enlarge the instability strips, such as the Coriolis force due to fast rotation \citep{Bouabid2013,Szewczuk2017} and radiative levitation due to atomic diffusion \citep{Deal2016,Rehm2024}. All of this makes comparing observational instability strips from surveys of pulsators with predicted strips based on just one choice of input physics of limited value. This was already highlighted from the  dominant frequencies found in the Gaia DR3 light curves of g-mode pulsators in \citet{DeRidder2023}, got stressed again in the review on asteroseismology of fast rotators by \citet{AertsTkachenko2024}, and is reinforced here from our indistinguishable TESS amplitude distributions for the $\gamma\,$Dor and SPB classes shown in Fig.\,\ref{fig:amp_dist}. It is for this reason that we merge the SPB and $\gamma$\,Dor classes in Sec.~\ref{sec:classification}.

\subsection{Gaia amplitude ratios}

\begin{figure}[t]
    \begin{center}
        \includegraphics[]{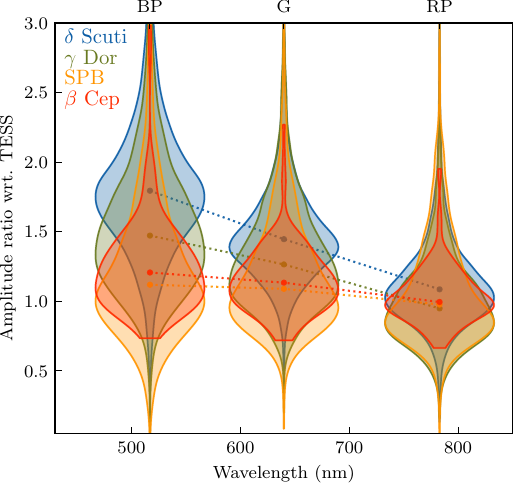}
    \end{center}
    \caption{\label{fig:amp_ratio} Violin plot of the amplitude ratios for the four samples deduced from the light curves in the three Gaia bandpasses with respect to the amplitude in the TESS band (covering approximately 600 - 1,000 nm). The shaded regions indicate the density distributions for each of the samples, with the solid points showing the median. For clarity, we do not add the density distribution centred around value 1.0 for the TESS bandpass itself.}
\end{figure}

So far we worked with the Gaia G passband. It covers the wavelengths between 330\,nm to 1050\,nm, with peak sensitivity at 640\,nm. But DR3 also delivered the light curves in the RP and BP colour bands. In practice, these BP and RP bands are blue and red cuts of the broad G band, covering the ranges from 330\,nm to 680\,nm (BP) and from 630\,nm to 1050\,nm (RP), with maximum responses at 517\,nm  and 783\,nm, respectively \citep{Weiler2018}. On the other hand, the TESS detector passband spans from 600\,nm to 1000\,nm with
central wavelength at 787\,nm \citep{Ricker2015}.

Having time series data with colour information is advantageous for asteroseismology in the case of ambiguous mode identification in terms of the spherical harmonic wavenumbers $(l,m)$ characterising the geometry of the mode. Indeed, the theoretical expression for the observed amplitudes of a mode described by the spherical harmonic function $Y_l^m$ and viewed upon at an inclination angle $i$ depends on the wavelength, via the perturbations
of the atmospheric flux and limb darkening
caused by the mode \citep[see Eq.\,(6.29) in ][]{Aerts2010}. The dependencies of this expression on the azimuthal order $m$ and on the inclination angle $i$ of the star's pulsation symmetry axis drop out of the expression of amplitude ratios for different wavelengths. This is why observed amplitude ratios deduced for light curves observed in different passbands have been used to identify the mode degree $l$ of main-sequence pulsators \citep[e.g.,][]{Heynderickx1994,Breger2000Scuti,Dupret2003Photometric,Aerts2004Longterm,DeCat2017Pulsating,Brunsden2018Frequency}. All these applications of mode identification assumed one fixed set of input physics for the theoretical predictions. We now know from space asteroseismology that this is not appropriate for such pulsators \citep{Aerts2021-RMP,Johnston2021,Pedersen2022}.

Although the Coriolis and centrifugal forces complicate this capacity of mode identification in fast rotators such as $\delta\,$Scuti stars \citep{Daszynska2002} and SPB stars \citep{Townsend2003}, we have a good understanding of how they do so. Therefore,
it was recently suggested by \citet{AertsTkachenko2024} to exploit amplitude ratios for stars whose identification of $(l,m)$ is already established. Indeed, in this case, any diversity in observed amplitude ratios encapsulates differences in the internal, atmospheric, and limb-darkening physics of the star. Figure\,11 in \citet{AertsTkachenko2024} illustrates this (future) potential opportunity from measured amplitude ratios of prograde dipole gravito-inertial modes observed in both Gaia and {\it Kepler\/} data of 23 $\gamma\,$Dor stars.
Here, we illustrate the potential of amplitude ratios from combined Gaia and TESS light curves for the four samples of new Gaia DR3 pulsators.

For all the stars with consistent dominant frequency and consistent classification (Section~\ref{sec:classification}) in the Gaia G and TESS passbands, we computed the ratios of the G, BP, and RP amplitudes of that frequency with respect to the amplitude in the TESS passband. We show a violin plot of the results for the four classes in Fig.\,\ref{fig:amp_ratio}. This figure is in line with expectations for low-degree ($l\leq 2$) mode behaviour in stars with slow to moderate rotation, whose ratios are predicted to decrease with increasing wavelength for the three Gaia passbands \citep[e.g.,][]{Heynderickx1994,DeRidder2004}. Comparison between Fig.\,11 in \citet{AertsTkachenko2024} and the violin plot in Fig.\,\ref{fig:amp_ratio} illustrates the superiority of TESS over {\it Kepler\/} for this type of exploratory research based on the dominant pulsation mode, given that the TESS pulsators are generally much brighter than the {\it Kepler\/} $\gamma\,$Dor stars, leading to more precise amplitude ratios. Given its potential for asteroseismology, we provide the Gaia amplitude ratios alongside the database of prewhitened modes as supplementary data.

\section{Re-classification of the pulsators from TESS}
\label{sec:classification}

\begin{figure}[t]
    \begin{center}
        \includegraphics[width=\linewidth]{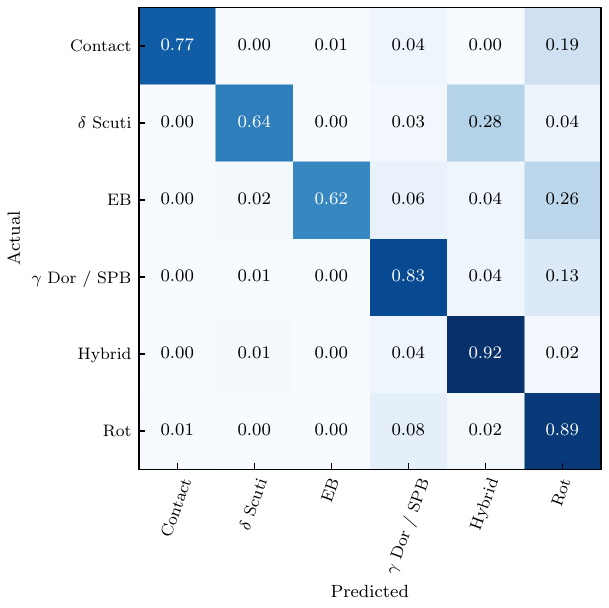}
    \end{center}
    \caption{\label{fig:confusionmatrix} Confusion matrix for the Random Forest classifier normalized against the true values. Here, hybrid refers to 
the simultaneous presence of $\delta$ Scuti 
p-mode pulsations and $\gamma$\,Dor or SPB g-mode pulsations.}
\end{figure}

\begin{table}[t]
    \caption[]{Classifications of the p- and g-mode candidate sample. The full table in electronic format, with probabilities for each class, is available online.}
    \label{tab:classification}

    \begin{tabular}{rrlr}
        \hline
        DR3 Source ID       & Sector & Class            & Probability \\
        \hline
        2070667440659268352 & 14     & $\delta$ Scuti   & 0.26        \\
        2247307763228746240 & 14     & Hybrid           & 0.67        \\
        5311721917688649856 & 10     & $\delta$ Scuti   & 0.35        \\
        5617799085534387968 & 7      & $\delta$ Scuti   & 0.67        \\
        5657691905702501888 & 9      & $\gamma$ Dor/SPB & 0.93        \\
                            & \vdots                                  \\
        2925330438953873024 & 7      & Hybrid           & 0.95        \\
        2164506978535115776 & 16     & Hybrid           & 0.41        \\
        5516571310575369472 & 8      & $\delta$ Scuti   & 0.74        \\
        5964132569560169472 & 12     & $\delta$ Scuti   & 0.57        \\
        5661189937524776320 & 9      & $\gamma$ Dor/SPB & 0.90        \\
        \hline
    \end{tabular}
\end{table}

We now re-evaluate the Gaia DR3 variability classification from \citet{DeRidder2023} by relying on the highly sampled and more precise TESS light curves.

\subsection{Training sample}

To distinguish between various classes of variability, in particular, intrinsic and extrinsic variability, we have implemented a simple feature-based Random Forest classifier, similar to \citet{Audenaert2021TESS} and \citet{Barbara2022Classifying}. This classifier seeks to identify different types of variability based on extracted singular value features of the light curve. These features, calculated with \textsc{pycatch22} \citep{Lubba2019Catch22}, are spread across different categories concerning linear autocorrelation and periodicity of the flux, extreme events, distributions that ignore time ordering, and more. These features have been chosen such that they are minimally redundant and capture the largest possible variance of the input time series.

Our training sample for the classifier is sourced from both {\it Kepler\/} and TESS. For {\it Kepler\/}, we use the sample compiled by \citet{Barbara2022Classifying}, which consists of high-level variability classifications pertaining to A/F-type stars, including contact and detached binaries, $\delta$ Scuti stars, $\gamma$ Dor stars, RR Lyrae variables, along with rotational and non-variables. Given that this sample relies heavily on data derived from the \textit{Kepler} mission, it poses certain challenges when applied to classify TESS data; the majority of the stars within the sample are of such faint magnitude that their variability signal cannot be observed within the TESS data, hence directly comparing the TESS light curves with the \textit{Kepler} labels is not feasible. As an alternative, we have devised a strategy where we modify the \textit{Kepler} light curves to mirror the single-sector observations of the TESS photometry. The modifications we have implemented on the \textit{Kepler} light curves are as follows; limiting the time span to a duration of less than 27 days, adding noise proportional to the magnitude of the star, and introducing a data gap at 13.7 days to simulate the TESS downlink.

We further construct a TESS training sample compiled against a series of A/F variability papers \citep{Skarka2022Periodic,Sikora2019a,Garcia2022b,Garcia2022a, Shi2023Catalog}, focusing on either the classification of A/F stars or targeting a specific variable type in TESS. The Skarka sample consists of variable A/F stars in the Northern continuous viewing zone, the Sikora sample contains rotationally variable A-type stars, the Garcia sample contains 60 $\gamma$\,Dor stars with a long observational baseline, and the Shi sample contains 286 SPB stars.

Each sample contains a slightly different type of classification, which we homogenize into new categories. In particular, we merge all the ellipsoidal and semi-detached binaries into the ``contact" class, leaving the eclipsing binary (EB) class for purely detached cases. We also merge the two hybrid classes ($\gamma$ Dor + $\delta$ Scuti vs.\ $\delta$ Scuti + $\gamma$ Dor) into a single hybrid class regardless of which variability type is more prominent. We also merge the SPB and $\gamma$\,Dor pulsators into a single class, 
as already motivated in the previous section (additional reasons are discussed below). The remaining classes are the pure $\delta$ Scuti pulsators and pure rotational variables (`Rot'). We discard the Skarka sample containing "VAR" sources -- stars deemed to be variable with an indeterminate classification.

The majority of the stars in this additional training sample are located in the TESS continuous viewing zone (CVZ), such that each star has multiple sectors of observations 
up to almost a year in length. On the other hand, our classification sample -- the candidate p- and g-mode pulsators -- are typically observed in only one or two sectors as a consequence of being distributed randomly across the sky. As a result, we do not stitch the light curves of any of the targets in the training sample. Instead, we compute their features on a per-sector basis and consider each sector of observations as a separate input. That is, a single target in the training sample can contribute to the final sample multiple times. We note that this will lead to larger ambiguity in the classification sample. For example, observations of true $\gamma$\,Dor pulsations might be unresolved in a single sector, such that they are confused with a rotational signal. Likewise, variables such as eclipsing or ellipsoidal variables may have variability periods which exceed the single-sector observations.

\subsection{Feature extraction and classification}

\begin{table}[t]
    \caption{Results of classification for each sample, showing the breakdown of individual classifications as a fraction of the total sample. The number in brackets represents the fraction of the sample for the class.}
    \begin{equation*}
        \begin{aligned}
            \text{$\gamma$ Dor (N=10,047)}   & \left\{
            \begin{array}{rl}
                \textrm{\textbf{$\gamma$ Dor / SPB}} & 6,489~(0.65) \\
                \textrm{Rotation}                    & 2,416~(0.24) \\
                \textrm{\textbf{Hybrid}}             & 618~(0.06)   \\
                \textrm{Eclipsing binary}            & 251~(0.02)   \\
                \textrm{Contact binary}              & 205~(0.02)   \\
                \textrm{$\delta$ Scuti}              & 58~(\sim0)   \\
            \end{array}\right. \\
            \text{$\delta$ Scuti (N=45,648)} & \left\{
            \begin{array}{rl}
                \textrm{\textbf{$\delta$ Scuti}} & 19,226~(0.42) \\
                \textrm{\textbf{Hybrid}}         & 15,395~(0.34) \\
                \textrm{Rotation}                & 4,371~(0.10)  \\
                \textrm{Eclipsing binary}        & 3,656~(0.08)  \\
                \textrm{$\gamma$ Dor / SPB}      & 2,962~(0.06)  \\
                \textrm{Contact binary}          & 38~(\sim0)    \\
            \end{array}\right.    \\
            \text{SPB (N=2,795)}             & \left\{
            \begin{array}{rl}
                \textrm{\textbf{$\gamma$ Dor / SPB}} & 1,481 ~(0.53) \\
                \textrm{Rotation}                    & 948~(0.34)    \\
                \textrm{Hybrid}                      & 209 ~(0.07)   \\
                \textrm{Eclipsing binary}            & 88 ~(0.03)    \\
                \textrm{Contact binary}              & 59 ~(0.02)    \\
                \textrm{$\delta$ Scuti}              & 10~(\sim0)    \\
            \end{array}\right.       \\
        \end{aligned}
    \end{equation*}
    \label{tab:class_compare}
\end{table}

We apply a uniform processing of each light curve prior to feature extraction. This processing includes applying a Gaussian high-pass filter to remove long-term trends and dividing the light curve by the standard deviation of its flux (Z-scoring) to ensure normality across the light curve sample. Similar to the training sample, each target is classified on a sector-by-sector basis, so that a single target can have multiple classifications across different sectors.

We use a greedy feature-selection algorithm to pick out a sample of 7 highly orthogonal features from the original 22 features calculated. These are the features that best discriminate amongst the variability classes. We also include several additional features we consider important to the classification which are not calculated in \textsc{pycatch22} but are known to help distinguish variability (see, e.g., \citealt{Murphy2019Gaiaderived}). These features are the skewness of the flux for discriminating eclipsing binaries, as well as the skewness of the amplitude spectrum at frequencies less than 1~d$^{-1}$, less than 5~d$^{-1}$, and greater than 5~d$^{-1}$. The frequency domain skewness indicators measure the effective power contained in different regions of the amplitude spectrum: $\delta$ Scuti variables will typically have significantly higher skewness at higher frequencies, and hybrid pulsators will have strong skewness in both regions. Finally, we include the dominant frequency and amplitude of pulsation (in Z-scored units), the Gaia BP-RP colour index, and the Gaia reduced unit weight error.

Figure~\ref{fig:confusionmatrix} shows the confusion matrix for our training and test data. For single-sector observations, the classifier appears accurate, especially for hybrid pulsators and rotational variables. Unsurprisingly, the $\delta$ Scuti class is strongly confused with the hybrid class since the training sample was mostly based on TESS data exceeding multiple sectors. As a result, not all modes in the hybrid pulsators are resolved in only a single sector. Curiously, the eclipsing binary class has some overlap with the rotational variable class. This is likely because the EB class consists of semi-detached (EA) and W Ursae Majoris (EW) binaries. Only the contact class contains ellipsoidal binaries. Finally, the $\gamma$\,Dor class is weakly confused with the rotational variables. Again, this is expected; a single sector of observations limits the ability to resolve modes, thus the single rotation peak can be mistaken for a $\gamma$ Dor pulsation and vice-versa.

We run the classifier on our sample of candidate $\delta$ Scuti and $\gamma$ Dor pulsators. Note that we do not classify the $\beta$ Cep sample which we instead manually inspect. The results of the classifier for the sample along with their class probability are presented in Table\,~\ref{tab:classification}, with the breakdown of each sample and resulting classification in Table\,~\ref{tab:class_compare}. The number of classified stars is slightly less than the number of available light curves discussed in Sec.~2.2 because some light curves are completely dominated by poor data and were discarded from the sample.

It is important to note that a single star observed in multiple sectors will have a classification for each sector in the table. For example, TIC 38458616 has been observed for 13 sectors in the first two years of TESS and has an independent classification per sector. Nine of the sectors predict it to be $\gamma$\,Dor variable, and four predict it to be a contact binary, with the majority of sectors having a low probability for their respective class. A closer inspection of the stitched light curve reveals the target to indeed be a binary system. While stitching individual light curves may give better results for poorly resolved modes, we strive to instead maintain a uniform input sample for the classifier. We show examples of high probability classifications in Fig.~\ref{example_class}.

Finally, we make an additional cut on the resulting classifications. To ensure that the rotational and eclipsing binary variables are reasonably well-separated from the g-mode pulsators, we apply a data cut such that the second-highest amplitude frequency can not be half of the dominant frequency within a tolerance of 0.01~d$^{-1}$ based on the prewhitening performed in Sec.~3.1. This is done because rotational variables and eclipsing binaries typically show a subharmonic frequency at half the dominant due to their strongly non-sinusoidal light curves. Indeed, for eclipsing binaries, this subharmonic is usually the true frequency. Making this additional cut removes 498 candidates from the $\gamma$ Dor sample.


\begin{figure*}[t]
    \begin{center}
        \includegraphics[]{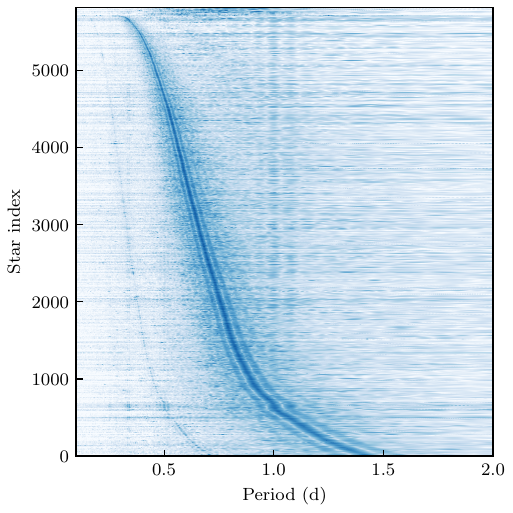}
        \includegraphics[]{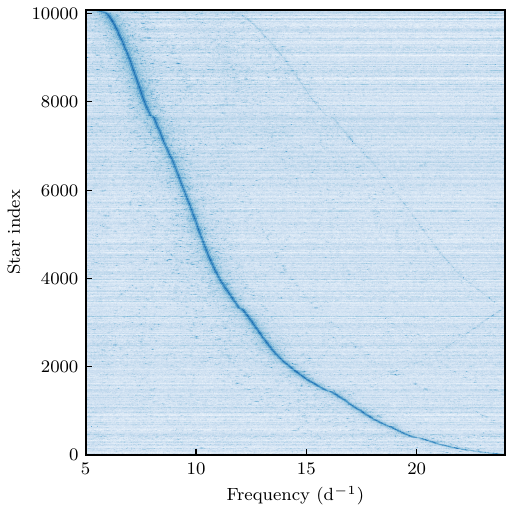}
    \end{center}
    \caption{\label{stacked_ps} Stacked amplitude spectra of the g-mode candidate sample (left, in period space) and $\delta$ Scuti candidate sample (right, in frequency space) for which the prediction probability is greater than 0.5. Each star forms a single row, sorted by the dominant pulsation, with color corresponding to amplitude. For the g-mode sample, which combines both $\gamma$\,Dor and SPB stars, we see a distinct secondary ridge associated with either a harmonic of the dominant frequency or the expected $\ell=2$ dipole modes seen in \citet{Li2020Gravitymode}. For the $\delta$ Scuti sample, we see ridges associated with the first and second overtones, as well as a harmonic line.
    }
\end{figure*}

\begin{figure}[t]
    \begin{center}
        \includegraphics[]{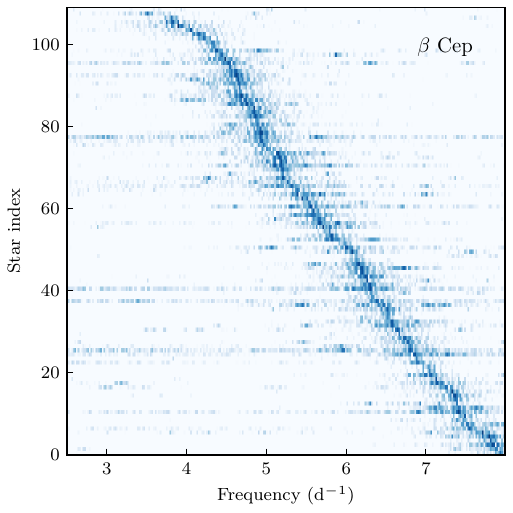}
    \end{center}
    \caption{\label{stacked_ps_bcep_spb} Stacked amplitude spectra of the $\beta$ Cep sample sorted by dominant pulsation frequency \rev{for those class members whose dominant mode occurs between 3 and 8\,d$^{-1}$}.}
\end{figure}

Using these classifications, we construct the stacked 
mplitude spectra. That is, we calculate the 
amplitude spectrum for each star and stack it according to the dominant pulsation frequency for the classified g-mode sample ($\gamma$ Dor / SPB), the $\delta$ Scuti stars, and the $\beta$ Cep stars, 
whose prediction probability is above 0.5.
We jointly visualize the $\gamma$ Dor and SPB sample on the same figure. We show the stacked 
amplitude spectra in Fig.~\ref{stacked_ps}. Each row displays the amplitude spectrum of one star sorted vertically by the dominant pulsation period. For the g-mode sample, we observe two distinct ridges, along with a third very faint ridge. The primary ridge is likely dominated by $\ell=1$ $m=1$ g-modes, with the secondary by 
either
lower amplitude $\ell=2$ modes similar to what was seen by \citet{GangLi2020} for the {\it Kepler\/} data,
or caused by the harmonic of the dominant mode.
We note that the highest amplitude ridge is likely not all pure dipole modes as the ridge is formed from sorting by the dominant period. Note also the presence of a purely vertical ridge and sidelobes at 1~d$^{-1}$ caused by the known TGLC systematic.

The stacked $\delta$ Scuti figure shows four obvious ridges, corresponding to the primary, first and second overtones, and harmonic frequency. These lines are known properties of $\delta$ Scuti stars, visualized more commonly in Peterson diagrams \citep{Netzel2022Frequency, Pietrukowicz202010}. Finally, the $\beta$ Cep sample shows no obvious ridges as expected for these low-order p- and g-mode pulsators \citep{Stankov2005}.

\subsection{Prioritised catalogue of new pulsators ranked by asteroseismic potential}
\label{sec:finaltable}

\begin{table*}
    \tabcolsep=3pt
    \caption{Rank ordered tables for the $\gamma$ Dor, $\delta$ Scuti, and SPB, and $\beta$ Cep classified pulsators. Note that the tables are separated according to their candidate and classification status -- g-mode pulsators are the g-mode candidates from \citet{Aerts2023Astrophysical}, and the p-mode pulsators are the candidates from \citet{DeRidder2023}. In the table, P and  N$_f$ refer to the probability of the classification and the number of independent modes respectively. Sectors is the number of non-contiguous sectors in which the target falls on TESS cameras calculated up to Cycle 6. The full version in electronic format is available online.}
    \label{tab:rank}
    \begin{tabular}{llllcrrrrrrrr}
        \hline
                  &                     &        &        & \textit{$\gamma$ Dor sample}                                                                          \\
        \hline
        TIC ID    & DR3 ID              & RA     & Dec    & Prediction                     & P    & N$_f$    & $f_1$    & $f_2$ & $A_1$ & $A_2$ & Sectors & Score \\
                  &                     & deg    & deg    &                                &      & d$^{-1}$ & d$^{-1}$ & ppt   & ppt   &       &                 \\
        \hline
        326258494 & 2249180128450868992 & 298.84 & 67.87  & GDOR/SPB                       & 0.98 & 17       & 2.62     & 2.24  & 12.21 & 0.72  & 37      & 0.79  \\
        259127682 & 2260889652408354944 & 291.59 & 67.19  & GDOR/SPB                       & 0.84 & 24       & 1.08     & 1.18  & 9.07  & 4.61  & 36      & 0.79  \\
        364588332 & 4648571270586910976 & 85.13  & -75.69 & GDOR/SPB                       & 0.91 & 22       & 1.31     & 1.30  & 9.71  & 5.38  & 35      & 0.79  \\
                  &                     &        &        & \vdots                                                                                                \\
        \hline
        \hline
                  &                     &        &        & \textit{$\delta$ Scuti sample}                                                                        \\
        \hline
        TIC ID    & DR3 ID              & RA     & Dec    & Prediction                     & P    & N$_f$    & $f_1$    & $f_2$ & $A_1$ & $A_2$ & Sectors & Score \\
        \hline
        233617727 & 2253706710446026496 & 284.02 & 64.80  & DSCT                           & 0.75 & 26       & 13.02    & 13.09 & 15.41 & 3.60  & 36      & 0.77  \\
        176960346 & 5266903384176544640 & 101.21 & -70.16 & DSCT                           & 0.60 & 37       & 6.54     & 10.97 & 14.79 & 2.83  & 33      & 0.77  \\
        38461275  & 4670142206953240448 & 61.21  & -64.17 & DSCT                           & 0.62 & 36       & 9.82     & 10.53 & 8.04  & 2.46  & 32      & 0.76  \\
                  &                     &        &        & \vdots                                                                                                \\
        \hline

        \hline
                  &                     &        &        & \textit{SPB sample}                                                                                   \\
        \hline
        TIC ID    & DR3 ID              & RA     & Dec    & Prediction                     & P    & N$_f$    & $f_1$    & $f_2$ & $A_1$ & $A_2$ & Sectors & Score \\
        \hline
        267543987 & 2264463198340787968 & 289.43 & 72.95  & GDOR/SPB                       & 0.62 & 16       & 1.28     & 1.00  & 20.90 & 0.57  & 36      & 0.66  \\
        349784439 & 5288831253807922560 & 113.20 & -62.78 & GDOR/SPB                       & 0.79 & 7        & 2.98     & 1.49  & 7.93  & 5.09  & 34      & 0.63  \\
        300382254 & 5269074232447890048 & 111.28 & -67.76 & GDOR/SPB                       & 0.64 & 14       & 1.24     & 0.62  & 16.69 & 0.80  & 34      & 0.63  \\
                  &                     &        &        & \vdots                                                                                                \\
        \hline

        \hline
                  &                     &        &        & \textit{$\beta$ Cep sample}                                                                           \\
        \hline
        TIC ID    & DR3 ID              & RA     & Dec    & Prediction                     & P    & N$_f$    & $f_1$    & $f_2$ & $A_1$ & $A_2$ & Sectors & Score \\
        \hline
        145594454 & 5328039318762759552 & 132.55 & -49.00 & ---                            & ---- & 40       & 5.71     & 5.80  & 15.24 & 8.36  & 6       & 0.49  \\
        276171115 & 2055651749653738112 & 303.31 & 34.02  & ---                            & ---  & 34       & 5.21     & 5.12  & 15.52 & 4.23  & 8       & 0.46  \\
        90964721  & 2055288395420325760 & 302.00 & 33.67  & ---                            & ---  & 34       & 4.06     & 3.69  & 12.94 & 1.54  & 8       & 0.46  \\
                  &                     &        &        & \vdots                                                                                                \\
        \hline
    \end{tabular}
\end{table*}

We present a catalogue of 
all the
new pulsators ranked by their asteroseismic potential (Table~\ref{tab:rank}). 
As an example for the $\gamma$ Dor variables, we quantify how likely they are to be true $\gamma$ Dor stars, and how viable we believe their pulsation modes are for typical g-mode analyses (i.e., \citealt{Li2020Gravitymode}). This table is a combination of the class prediction probability, the number of currently available sectors (calculated using \textsc{tess-point}; \citealt{2020ascl.soft03001B}), and cuts made based on prewhitening and combination frequency removal. We weigh the contribution to the `score' equally with the following criteria; 1. The number of sectors that the target falls on a TESS camera up to Cycle 6 of TESS observations divided by the maximum number of sectors possible for a CVZ target. 2. The predicted class probability, calculated as the mean probability of each sector. For targets with multiple sectors of observation, we find the mean probability of each predicted class and choose the class with the highest probability. Since the $\beta$ Cep sample has no predictions, we do not include this in their scoring. 3. the number of observed independent modes found during the prewhitening process after removal of harmonics and combination frequencies.

We include all columns used to calculate this score for users wishing to prioritize 
follow-up studies or work with alternative features, and show a few high-probability classifications in Fig.~\ref{example_class}.

\rev{Finally, we briefly comment on the overlap of our asteroseismic sample with the sample provided by IJSpeert et al., 2024 (submitted). Their sample consists of 14,573 characterized eclipsing binary systems observed by TESS across the O- to F-type stars. Due to differing selection methods, our overlap consists of only 339 stars. However, a comparison of dominant and secondary frequencies indicates good agreement between the two samples.}

\section{Conclusions}

In this paper we have examined the pulsators of spectral type
O, B, A, or F classified from Gaia DR3 photometry by \citet{DeRidder2023} and having
TESS high-cadence light curves. A comparison of dominant frequencies present in these independent light curves indicates that Gaia is extremely good at measuring g-mode frequencies (approximately 80\% precision when compared against TESS), and reasonably effective at higher frequencies (69\%). We note that for the higher frequency p-modes, it is unclear whether the Gaia or TESS data is more accurate for measuring the `true' dominant frequency. The 30-minute cadence of the TESS data precludes accurate measurement of signals above 24~d$^{-1}$, whereas the Gaia photometry suffers from instrumental effects at mmag level and its unequal sampling means there is no clearly defined Nyquist limit. As such, we consider the 69\% precision for the p-modes to be a worst-case scenario.

A comparison of amplitudes for the dominant and secondary frequencies for each variable class reveals that $\gamma$ Dor and SPB variables have indistinguishable amplitude distributions. Prior work on variability classification supports this result. While colour information has been used to distinguish between the two classes of pulsators on the basis of instability computations, 
\citet{DeRidder2023} and 
\citet{Aerts2023Astrophysical} found there to be a `continuum' of g-mode pulsators bridging the predicted strips. Said differently, both Gaia DR3 and TESS reveal that g-mode pulsators occur along the main sequence covering an effective temperature ranging from roughly 6500\,K to 18\,000\,K. According to instability computations available in the literature, a fraction of these g-mode pulsators are too cold to be SPB stars and too hot to be $\gamma\,$Dor stars. The large ranges in  
effective temperature and luminosity for the Gaia DR3 $\gamma\,$Dor and SPB stars discussed in \citet{Aerts2023Astrophysical} and now confirmed with TESS  point to a  lack of physical ingredients 
in excitation predictions, 
such as rotation, radiative levitation, nonlinear mode coupling (and tidal mode excitation not discussed here). The combined Gaia DR3 and TESS light curves make us conclude that there is one large global region of g-mode pulsations excited along the main sequence, which is likely caused by a multitude of non-exclusive physical phenomena. This suggestion from \citet{Aerts2023Astrophysical} is now solidified 
here from our confirmation of the nature of these g-mode pulsators in our catalogue, thanks to their TESS light curves. We also note that although g-modes appear to be found across that entire temperature range, not all stars pulsate in g-modes and not all pulsators are hybrids.

The classification of the TESS light curves reveals that the original Gaia variability classification done by Coordination Unit 7 of the mission 
\citep[see ][]{Holl2018,Eyer2019,Eyer2023,Rimoldi2023,DeRidder2023}
is remarkably accurate. For each candidate variable from Gaia, we find that the majority of their TESS light curve classifications are in good agreement with their Gaia classification. These results point to around 6,000 new $\gamma$\,Dor, 20,000 new $\delta$\,Scuti, and 1,481 new SPB pulsators confirmed by TESS.
While the TESS light curve classification is expected to be more accurate than Gaia, we note that it is not perfect. Notably, the low-frequency g-mode pulsators are easily confused with rotational variables.

We have made available several tables and datasets from the results of this paper, including; prewhitened frequencies, amplitudes (in Gaia and TESS), phases, and false alarm levels to 1\% significance level for every target, classifications and their respective probabilities for each sector of observation, and a rank-ordered table of useful candidate pulsators. It is our hope that the results presented here will enable future ensemble asteroseismic studies of hot non-radial pulsators \rev{on the main sequence}, especially with the release of Gaia DR4 and DR5, as well as with the upcoming PLATO mission \citep{Rauer2024}.

\begin{acknowledgements}
    The authors thank Timothy Bedding for helpful comments and discussions.
    The research leading to these results has received funding from the KU\,Leuven Research Council (grant C16/18/005: PARADISE) and from the European Research
    Council (ERC) under the Horizon Europe programme (Synergy Grant
    agreement N$^\circ$101071505: 4D-STAR). While partially funded by the European Union, views and opinions expressed are however those of the author(s) only and do not necessarily reflect those of the European Union or the European Research Council. Neither the European Union nor the granting authority can be held responsible for them. This research was supported by the Munich Institute for Astro-, Particle and BioPhysics (MIAPbP) which is funded by the Deutsche Forschungsgemeinschaft (DFG, German Research Foundation) under Germany´s Excellence Strategy – EXC-2094 – 390783311. 
    The MIAPbP research program facilitated the interactions between the two authors, which ultimately led to the current paper.
\end{acknowledgements}

\newpage
\bibliographystyle{aa} 
\bibliography{Hey-Aerts.bib, library.bib} 

\begin{thebibliography}{}
\expandafter\ifx\csname natexlab\endcsname\relax\def\natexlab#1{#1}\fi
\providecommand{\url}[1]{\href{#1}{#1}}
\providecommand{\dodoi}[1]{doi:~\href{http://doi.org/#1}{\nolinkurl{#1}}}
\providecommand{\doeprint}[1]{\href{http://ascl.net/#1}{\nolinkurl{http://ascl.net/#1}}}
\providecommand{\doarXiv}[1]{\href{https://arxiv.org/abs/#1}{\nolinkurl{https://arxiv.org/abs/#1}}}

\bibitem[{{Aerts}(2021)}]{Aerts2021-RMP}
{Aerts}, C. 2021, Reviews of Modern Physics, 93, 015001,
  \dodoi{10.1103/RevModPhys.93.015001}

\bibitem[{{Aerts} {et~al.}(2010){Aerts}, {Christensen-Dalsgaard}, \&
  {Kurtz}}]{Aerts2010}
{Aerts}, C., {Christensen-Dalsgaard}, J., \& {Kurtz}, D.~W. 2010,
  {Asteroseismology, Springer-Verlag Heidelberg}

\bibitem[{Aerts {et~al.}(2004)Aerts, Cuypers, De~Cat, Dupret, De~Ridder, Eyer,
  Scuflaire, \& Waelkens}]{Aerts2004Longterm}
Aerts, C., Cuypers, J., De~Cat, P., {et~al.} 2004, Astronomy and Astrophysics,
  415, 1079, \dodoi{10.1051/0004-6361:20034628}

\bibitem[{Aerts {et~al.}(2023)Aerts, Molenberghs, \&
  De~Ridder}]{Aerts2023Astrophysical}
Aerts, C., Molenberghs, G., \& De~Ridder, J. 2023, Astrophysical Properties of
  15062 {{Gaia DR3}} Gravity-Mode Pulsators: Pulsation Amplitudes, Rotation,
  and Spectral Line Broadening,  {arXiv}, \dodoi{10.48550/arXiv.2302.07870}

\bibitem[{{Aerts} \& {Rogers}(2015)}]{Aerts2015}
{Aerts}, C., \& {Rogers}, T.~M. 2015, \apjl, 806, L33,
  \dodoi{10.1088/2041-8205/806/2/L33}

\bibitem[{{Aerts} \& {Tkachenko}(2024)}]{AertsTkachenko2024}
{Aerts}, C., \& {Tkachenko}, A. 2024, \aap, in press, arXiv:2311.08453,
  \dodoi{10.48550/arXiv.2311.08453}

\bibitem[{{Aerts} {et~al.}(2018){Aerts}, {Molenberghs}, {Michielsen},
  {Pedersen}, {Bj{\"o}rklund}, {Johnston}, {Mombarg}, {Bowman}, {Buysschaert},
  {P{\'a}pics}, {Sekaran}, {Sundqvist}, {Tkachenko}, {Truyaert}, {Van Reeth},
  \& {Vermeyen}}]{Aerts2018}
{Aerts}, C., {Molenberghs}, G., {Michielsen}, M., {et~al.} 2018, \apjs, 237,
  15, \dodoi{10.3847/1538-4365/aaccfb}

\bibitem[{{Antoci} {et~al.}(2019){Antoci}, {Cunha}, {Bowman}, {Murphy},
  {Kurtz}, {Bedding}, {Borre}, {Christophe}, {Daszy{\'n}ska-Daszkiewicz},
  {Fox-Machado}, {Garc{\'\i}a Hern{\'a}ndez}, {Ghasemi}, {Handberg}, {Hansen},
  {Hasanzadeh}, {Houdek}, {Johnston}, {Justesen}, {Kahraman Alicavus},
  {Kotysz}, {Latham}, {Matthews}, {M{\o}nster}, {Niemczura}, {Paunzen},
  {S{\'a}nchez Arias}, {Pigulski}, {Pepper}, {Richey-Yowell}, {Safari},
  {Seager}, {Smalley}, {Shutt}, {S{\'o}dor}, {Su{\'a}rez}, {Tkachenko}, {Wu},
  {Zwintz}, {Barcel{\'o} Forteza}, {Brunsden}, {Bogn{\'a}r}, {Buzasi},
  {Chowdhury}, {De Cat}, {Evans}, {Guo}, {Guzik}, {Jevtic}, {Lampens}, {Lares
  Martiz}, {Lovekin}, {Li}, {Mirouh}, {Mkrtichian}, {Monteiro}, {Nemec},
  {Ouazzani}, {Pascual-Granado}, {Reese}, {Rieutord}, {Rodon}, {Skarka},
  {Sowicka}, {Stateva}, {Szab{\'o}}, \& {Weiss}}]{Antoci2019}
{Antoci}, V., {Cunha}, M.~S., {Bowman}, D.~M., {et~al.} 2019, \mnras, 490,
  4040, \dodoi{10.1093/mnras/stz2787}

\bibitem[{Audenaert {et~al.}(2021)Audenaert, Kuszlewicz, Handberg, Tkachenko,
  Armstrong, Hon, Kgoadi, Lund, Bell, Bugnet, Bowman, Johnston, Garc{\'i}a,
  Stello, Moln{\'a}r, Plachy, Buzasi, Aerts, \&
  {collaboration}}]{Audenaert2021TESS}
Audenaert, J., Kuszlewicz, J.~S., Handberg, R., {et~al.} 2021, The Astronomical
  Journal, 162, 209, \dodoi{10.3847/1538-3881/ac166a}

\bibitem[{{Balona}(2015)}]{Balona2015}
{Balona}, L.~A. 2015, \mnras, 447, 2714, \dodoi{10.1093/mnras/stu2651}

\bibitem[{{Balona}(2016)}]{Balona2016}
---. 2016, \mnras, 457, 3724, \dodoi{10.1093/mnras/stw244}

\bibitem[{{Balona}(2024)}]{Balona2024}
---. 2024, \apj, submitted, arXiv:2310.09805, \dodoi{10.48550/arXiv.2310.09805}

\bibitem[{{Balona} \& {Dziembowski}(2011)}]{Balona2011-DSCT}
{Balona}, L.~A., \& {Dziembowski}, W.~A. 2011, \mnras, 417, 591,
  \dodoi{10.1111/j.1365-2966.2011.19301.x}

\bibitem[{{Balona} {et~al.}(2011{\natexlab{a}}){Balona}, {Cunha}, {Kurtz},
  {Brand{\~a}o}, {Gruberbauer}, {Saio}, {{\"O}stensen}, {Elkin}, {Borucki},
  {Christensen-Dalsgaard}, {Kjeldsen}, {Koch}, \& {Bryson}}]{Balona2011-Ap}
{Balona}, L.~A., {Cunha}, M.~S., {Kurtz}, D.~W., {et~al.} 2011{\natexlab{a}},
  \mnras, 410, 517, \dodoi{10.1111/j.1365-2966.2010.17461.x}

\bibitem[{{Balona} {et~al.}(2011{\natexlab{b}}){Balona}, {Pigulski}, {De Cat},
  {Handler}, {Guti{\'e}rrez-Soto}, {Engelbrecht}, {Frescura}, {Briquet},
  {Cuypers}, {Daszy{\'n}ska-Daszkiewicz}, {Degroote}, {Dukes}, {Garcia},
  {Green}, {Heber}, {Kawaler}, {Lehmann}, {Leroy}, {Molenda-{\.Z}aaowicz},
  {Neiner}, {Noels}, {Nuspl}, {{\O}stensen}, {Pricopi}, {Roxburgh}, {Salmon},
  {Smith}, {Su{\'a}rez}, {Suran}, {Szab{\'o}}, {Uytterhoeven},
  {Christensen-Dalsgaard}, {Kjeldsen}, {Caldwell}, {Girouard}, \&
  {Sanderfer}}]{Balona2011-Btype}
{Balona}, L.~A., {Pigulski}, A., {De Cat}, P., {et~al.} 2011{\natexlab{b}},
  \mnras, 413, 2403, \dodoi{10.1111/j.1365-2966.2011.18311.x}

\bibitem[{{Balona} {et~al.}(2019){Balona}, {Handler}, {Chowdhury}, {Ozuyar},
  {Engelbrecht}, {Mirouh}, {Wade}, {David-Uraz}, \& {Cantiello}}]{Balona2019}
{Balona}, L.~A., {Handler}, G., {Chowdhury}, S., {et~al.} 2019, \mnras, 485,
  3457, \dodoi{10.1093/mnras/stz586}

\bibitem[{Barbara {et~al.}(2022)Barbara, Bedding, Fulcher, Murphy, \&
  Van~Reeth}]{Barbara2022Classifying}
Barbara, N.~H., Bedding, T.~R., Fulcher, B.~D., Murphy, S.~J., \& Van~Reeth, T.
  2022, Monthly Notices of the Royal Astronomical Society, 514, 2793,
  \dodoi{10.1093/mnras/stac1515}

\bibitem[{{Bedding} {et~al.}(2020){Bedding}, {Murphy}, {Hey}, {Huber}, {Li},
  {Smalley}, {Stello}, {White}, {Ball}, {Chaplin}, {Colman}, {Fuller},
  {Gaidos}, {Harbeck}, {Hermes}, {Holdsworth}, {Li}, {Li}, {Mann}, {Reese},
  {Sekaran}, {Yu}, {Antoci}, {Bergmann}, {Brown}, {Howard}, {Ireland},
  {Isaacson}, {Jenkins}, {Kjeldsen}, {McCully}, {Rabus}, {Rains}, {Ricker},
  {Tinney}, \& {Vanderspek}}]{Bedding2020}
{Bedding}, T.~R., {Murphy}, S.~J., {Hey}, D.~R., {et~al.} 2020, \nat, 581, 147,
  \dodoi{10.1038/s41586-020-2226-8}

\bibitem[{{Bedding} {et~al.}(2023){Bedding}, {Murphy}, {Crawford}, {Hey},
  {Huber}, {Kjeldsen}, {Li}, {Mann}, {Torres}, {White}, \&
  {Zhou}}]{Bedding2023}
{Bedding}, T.~R., {Murphy}, S.~J., {Crawford}, C., {et~al.} 2023, \apjl, 946,
  L10, \dodoi{10.3847/2041-8213/acc17a}

\bibitem[{Bell {et~al.}(2017)Bell, Hermes, Vanderbosch, Montgomery, Winget,
  Dennihy, Fuchs, \& Tremblay}]{Bell2017Destroying}
Bell, K.~J., Hermes, J.~J., Vanderbosch, Z., {et~al.} 2017, The Astrophysical
  Journal, 851, 24, \dodoi{10.3847/1538-4357/aa9702}

\bibitem[{{Bouabid} {et~al.}(2013){Bouabid}, {Dupret}, {Salmon},
  {Montalb{\'a}n}, {Miglio}, \& {Noels}}]{Bouabid2013}
{Bouabid}, M.~P., {Dupret}, M.~A., {Salmon}, S., {et~al.} 2013, \mnras, 429,
  2500, \dodoi{10.1093/mnras/sts517}

\bibitem[{{Bowman}(2020)}]{Bowman2020-FrASS}
{Bowman}, D.~M. 2020, Frontiers in Astronomy and Space Sciences, 7, 70,
  \dodoi{10.3389/fspas.2020.578584}

\bibitem[{{Bowman} {et~al.}(2020){Bowman}, {Burssens}, {Sim{\'o}n-D{\'\i}az},
  {Edelmann}, {Rogers}, {Horst}, {R{\"o}pke}, \& {Aerts}}]{Bowman2020}
{Bowman}, D.~M., {Burssens}, S., {Sim{\'o}n-D{\'\i}az}, S., {et~al.} 2020,
  \aap, 640, A36, \dodoi{10.1051/0004-6361/202038224}

\bibitem[{{Bowman} {et~al.}(2018){Bowman}, {Buysschaert}, {Neiner},
  {P{\'a}pics}, {Oksala}, \& {Aerts}}]{Bowman2018}
{Bowman}, D.~M., {Buysschaert}, B., {Neiner}, C., {et~al.} 2018, \aap, 616,
  A77, \dodoi{10.1051/0004-6361/201833037}

\bibitem[{Bowman {et~al.}(2016)Bowman, Kurtz, Breger, Murphy, \&
  Holdsworth}]{Bowman2016Amplitude}
Bowman, D.~M., Kurtz, D.~W., Breger, M., Murphy, S.~J., \& Holdsworth, D.~L.
  2016, Monthly Notices of the Royal Astronomical Society, 460, 1970,
  \dodoi{10.1093/mnras/stw1153}

\bibitem[{{Bowman} {et~al.}(2019){Bowman}, {Burssens}, {Pedersen}, {Johnston},
  {Aerts}, {Buysschaert}, {Michielsen}, {Tkachenko}, {Rogers}, {Edelmann},
  {Ratnasingam}, {Sim{\'o}n-D{\'\i}az}, {Castro}, {Moravveji}, {Pope}, {White},
  \& {De Cat}}]{Bowman2019}
{Bowman}, D.~M., {Burssens}, S., {Pedersen}, M.~G., {et~al.} 2019, Nature
  Astronomy, 3, 760, \dodoi{10.1038/s41550-019-0768-1}

\bibitem[{Breger(2000)}]{Breger2000Scuti}
Breger, M. 2000, 210, 3

\bibitem[{{Briquet} {et~al.}(2007){Briquet}, {Hubrig}, {De Cat}, {Aerts},
  {North}, \& {Sch{\"o}ller}}]{Briquet2007}
{Briquet}, M., {Hubrig}, S., {De Cat}, P., {et~al.} 2007, \aap, 466, 269,
  \dodoi{10.1051/0004-6361:20066940}

\bibitem[{Brunsden {et~al.}(2018)Brunsden, Pollard, Wright, De~Cat, \&
  Cottrell}]{Brunsden2018Frequency}
Brunsden, E., Pollard, K.~R., Wright, D.~J., De~Cat, P., \& Cottrell, P.~L.
  2018, Monthly Notices of the Royal Astronomical Society, 475, 3813,
  \dodoi{10.1093/mnras/sty034}

\bibitem[{{Burke} {et~al.}(2020){Burke}, {Levine}, {Fausnaugh}, {Vanderspek},
  {Barclay}, {Libby-Roberts}, {Morris}, {Sipocz}, {Owens}, {Feinstein}, \&
  {Camacho}}]{2020ascl.soft03001B}
{Burke}, C.~J., {Levine}, A., {Fausnaugh}, M., {et~al.} 2020, {TESS-Point: High
  precision TESS pointing tool}, Astrophysics Source Code Library.
\newblock \doeprint{2003.001}

\bibitem[{{Burssens} {et~al.}(2019){Burssens}, {Bowman}, {Aerts}, {Pedersen},
  {Moravveji}, \& {Buysschaert}}]{Burssens2019}
{Burssens}, S., {Bowman}, D.~M., {Aerts}, C., {et~al.} 2019, \mnras, 489, 1304,
  \dodoi{10.1093/mnras/stz2165}

\bibitem[{{Burssens} {et~al.}(2020){Burssens}, {Sim{\'o}n-D{\'\i}az}, {Bowman},
  {Holgado}, {Michielsen}, {de Burgos}, {Castro}, {Barb{\'a}}, \&
  {Aerts}}]{Burssens2020}
{Burssens}, S., {Sim{\'o}n-D{\'\i}az}, S., {Bowman}, D.~M., {et~al.} 2020,
  \aap, 639, A81, \dodoi{10.1051/0004-6361/202037700}

\bibitem[{Burssens {et~al.}(2023)Burssens, Bowman, Michielsen,
  {Sim{\'o}n-D{\'i}az}, Aerts, Vanlaer, Banyard, Nardetto, Townsend, Handler,
  Mombarg, Vanderspek, \& Ricker}]{Burssens2023Calibration}
Burssens, S., Bowman, D.~M., Michielsen, M., {et~al.} 2023, Nature Astronomy,
  7, 913, \dodoi{10.1038/s41550-023-01978-y}

\bibitem[{{Cantiello} \& {Braithwaite}(2019)}]{Cantiello2019}
{Cantiello}, M., \& {Braithwaite}, J. 2019, \apj, 883, 106,
  \dodoi{10.3847/1538-4357/ab3924}

\bibitem[{{Cantiello} {et~al.}(2021){Cantiello}, {Lecoanet}, {Jermyn}, \&
  {Grassitelli}}]{Cantiello2021}
{Cantiello}, M., {Lecoanet}, D., {Jermyn}, A.~S., \& {Grassitelli}, L. 2021,
  \apj, 915, 112, \dodoi{10.3847/1538-4357/ac03b0}

\bibitem[{{C{\'o}rsico} {et~al.}(2019){C{\'o}rsico}, {Althaus}, {Miller
  Bertolami}, \& {Kepler}}]{Corsico2019}
{C{\'o}rsico}, A.~H., {Althaus}, L.~G., {Miller Bertolami}, M.~M., \& {Kepler},
  S.~O. 2019, \aapr, 27, 7, \dodoi{10.1007/s00159-019-0118-4}

\bibitem[{{Daszy{\'n}ska-Daszkiewicz}
  {et~al.}(2002){Daszy{\'n}ska-Daszkiewicz}, {Dziembowski}, {Pamyatnykh}, \&
  {Goupil}}]{Daszynska2002}
{Daszy{\'n}ska-Daszkiewicz}, J., {Dziembowski}, W.~A., {Pamyatnykh}, A.~A., \&
  {Goupil}, M.~J. 2002, \aap, 392, 151, \dodoi{10.1051/0004-6361:20020911}

\bibitem[{{David-Uraz} {et~al.}(2019){David-Uraz}, {Neiner}, {Sikora},
  {Bowman}, {Petit}, {Chowdhury}, {Handler}, {Pergeorelis}, {Cantiello},
  {Cohen}, {Erba}, {Keszthelyi}, {Khalack}, {Kobzar}, {Kochukhov},
  {Labadie-Bartz}, {Lovekin}, {MacInnis}, {Owocki}, {Pablo}, {Shultz},
  {ud-Doula}, \& {Wade}}]{David-Uraz2019}
{David-Uraz}, A., {Neiner}, C., {Sikora}, J., {et~al.} 2019, \mnras, 487, 304,
  \dodoi{10.1093/mnras/stz1181}

\bibitem[{De~Cat(2017)}]{DeCat2017Pulsating}
De~Cat, P. 2017, 152, 04001, \dodoi{10.1051/epjconf/201715204001}

\bibitem[{{De Ridder} {et~al.}(2004){De Ridder}, {Telting}, {Balona},
  {Handler}, {Briquet}, {Daszy{\'n}ska-Daszkiewicz}, {Lefever}, {Korn},
  {Heiter}, \& {Aerts}}]{DeRidder2004}
{De Ridder}, J., {Telting}, J.~H., {Balona}, L.~A., {et~al.} 2004, \mnras, 351,
  324, \dodoi{10.1111/j.1365-2966.2004.07791.x}

\bibitem[{{Deal} {et~al.}(2016){Deal}, {Richard}, \& {Vauclair}}]{Deal2016}
{Deal}, M., {Richard}, O., \& {Vauclair}, S. 2016, \aap, 589, A140,
  \dodoi{10.1051/0004-6361/201628180}

\bibitem[{Dupret {et~al.}(2005{\natexlab{a}})Dupret, Grigahc{\`e}ne, Garrido,
  De~Ridder, Scuflaire, \& Gabriel}]{Dupret2005Timedependent}
Dupret, M.-A., Grigahc{\`e}ne, A., Garrido, R., {et~al.} 2005{\natexlab{a}},
  Monthly Notices of the Royal Astronomical Society, 360, 1143,
  \dodoi{10.1111/j.1365-2966.2005.09114.x}

\bibitem[{Dupret {et~al.}(2005{\natexlab{b}})Dupret, Grigahc{\`e}ne, Garrido,
  Gabriel, \& Scuflaire}]{Dupret2005Convectionpulsation}
Dupret, M.-A., Grigahc{\`e}ne, A., Garrido, R., Gabriel, M., \& Scuflaire, R.
  2005{\natexlab{b}}, Astronomy \& Astrophysics, 435, 927,
  \dodoi{10.1051/0004-6361:20041817}

\bibitem[{Dupret {et~al.}(2003)Dupret, Ridder, Cat, Aerts, Scuflaire, Noels, \&
  Thoul}]{Dupret2003Photometric}
Dupret, M.-A., Ridder, J.~D., Cat, P.~D., {et~al.} 2003, Astronomy \&
  Astrophysics, 398, 677, \dodoi{10.1051/0004-6361:20021679}

\bibitem[{{Eyer} {et~al.}(2023){Eyer}, {Audard}, {Holl}, {Rimoldini},
  {Carnerero}, {Clementini}, {De Ridder}, {Distefano}, {Evans}, {Gavras},
  {Gomel}, {Lebzelter}, {Marton}, {Mowlavi}, {Panahi}, {Ripepi}, {Wyrzykowski},
  {Nienartowicz}, {Jevardat de Fombelle}, {Lecoeur-Taibi}, {Rohrbasser},
  {Riello}, {Garc{\'\i}a-Lario}, {Lanzafame}, {Mazeh}, {Raiteri}, {Zucker},
  {{\'A}brah{\'a}m}, {Aerts}, {Aguado}, {Anderson}, {Bashi}, {Binnenfeld},
  {Faigler}, {Garofalo}, {Karbevska}, {K{\'o}sp{\'a}l}, {Kruszy{\'n}ska},
  {Kun}, {Lanza}, {Leccia}, {Marconi}, {Messina}, {Molinaro}, {Moln{\'a}r},
  {Muraveva}, {Musella}, {Nagy}, {Pagano}, {Palaversa}, {Plachy}, {Pr{\v{s}}a},
  {Rybicki}, {Shahaf}, {Szabados}, {Szegedi-Elek}, {Trabucchi}, {Barblan},
  {Grenon}, {Roelens}, \& {S{\"u}veges}}]{Eyer2023}
{Eyer}, L., {Audard}, M., {Holl}, B., {et~al.} 2023, \aap, 674, A13,
  \dodoi{10.1051/0004-6361/202244242}

\bibitem[{Eze \& Handler(2024)}]{Eze2024Beta}
Eze, C.~I., \& Handler, G. 2024, \${\textbackslash}beta\$ {{Cephei}} Pulsators
  in Eclipsing Binaries Observed with {{TESS}},
  \dodoi{10.48550/arXiv.2403.12281}

\bibitem[{{Fritzewski} {et~al.}(2024){Fritzewski}, {Van Reeth}, {Aerts}, {Van
  Beeck}, {Gossage}, \& {Li}}]{Fritzewski2024}
{Fritzewski}, D.~J., {Van Reeth}, T., {Aerts}, C., {et~al.} 2024, \aap, 681,
  A13, \dodoi{10.1051/0004-6361/202347618}

\bibitem[{{Gaia Collaboration} {et~al.}(2018){Gaia Collaboration}, {Brown},
  {Vallenari}, {Prusti}, {de Bruijne}, {Babusiaux}, \&
  {Bailer-Jones}}]{Brown2018}
{Gaia Collaboration}, {Brown}, A.~G.~A., {Vallenari}, A., {et~al.} 2018, \aap,
  616, A1, \dodoi{10.1051/0004-6361/201833051}

\bibitem[{{Gaia Collaboration} {et~al.}(2016{\natexlab{a}}){Gaia
  Collaboration}, {Brown}, {Vallenari}, {Prusti}, \& {de Bruijne}}]{Brown2016}
{Gaia Collaboration}, {Brown}, A.~G.~A., {Vallenari}, A., {Prusti}, T., \& {de
  Bruijne}, J.~H.~J. e.~a. 2016{\natexlab{a}}, \aap, 595, A2,
  \dodoi{10.1051/0004-6361/201629512}

\bibitem[{{Gaia Collaboration} {et~al.}(2023{\natexlab{a}}){Gaia
  Collaboration}, {De Ridder}, {Ripepi}, {Aerts}, {Palaversa}, \&
  {Eyer}}]{DeRidder2023}
{Gaia Collaboration}, {De Ridder}, J., {Ripepi}, V., {et~al.}
  2023{\natexlab{a}}, \aap, 674, A36, \dodoi{10.1051/0004-6361/202243767}

\bibitem[{{Gaia Collaboration} {et~al.}(2016{\natexlab{b}}){Gaia
  Collaboration}, {Prusti}, {de Bruijne}, {Brown}, \& {Vallenari}}]{Prusti2016}
{Gaia Collaboration}, {Prusti}, T., {de Bruijne}, J.~H.~J., {Brown}, A.~G.~A.,
  \& {Vallenari}, A. e.~a. 2016{\natexlab{b}}, \aap, 595, A1,
  \dodoi{10.1051/0004-6361/201629272}

\bibitem[{{Gaia Collaboration} {et~al.}(2023{\natexlab{b}}){Gaia
  Collaboration}, {Vallenari}, {Brown}, {Prusti}, \& {de
  Bruijne}}]{Vallenari2023}
{Gaia Collaboration}, {Vallenari}, A., {Brown}, A.~G.~A., {Prusti}, T., \& {de
  Bruijne}, J.~H.~J. e.~a. 2023{\natexlab{b}}, \aap, 674, A1,
  \dodoi{10.1051/0004-6361/202243940}

\bibitem[{{Gaia Collaboration} {et~al.}(2019){Gaia Collaboration}, {Eyer},
  {Rimoldini}, {Audard}, {Anderson}, {Nienartowicz}, {Glass}, {Marchal},
  {Grenon}, {Mowlavi}, {Holl}, {Clementini}, {Aerts}, \& {Zwitter}}]{Eyer2019}
{Gaia Collaboration}, {Eyer}, L., {Rimoldini}, L., {et~al.} 2019, \aap, 623,
  A110, \dodoi{10.1051/0004-6361/201833304}

\bibitem[{{Garc{\'\i}a} \& {Ballot}(2019)}]{GarciaBallot2019}
{Garc{\'\i}a}, R.~A., \& {Ballot}, J. 2019, Living Reviews in Solar Physics,
  16, 4, \dodoi{10.1007/s41116-019-0020-1}

\bibitem[{{Garcia} {et~al.}(2022{\natexlab{a}}){Garcia}, {Van Reeth}, {De
  Ridder}, \& {Aerts}}]{Garcia2022b}
{Garcia}, S., {Van Reeth}, T., {De Ridder}, J., \& {Aerts}, C.
  2022{\natexlab{a}}, \aap, 668, A137, \dodoi{10.1051/0004-6361/202244365}

\bibitem[{{Garcia} {et~al.}(2022{\natexlab{b}}){Garcia}, {Van Reeth}, {De
  Ridder}, {Tkachenko}, {IJspeert}, \& {Aerts}}]{Garcia2022a}
{Garcia}, S., {Van Reeth}, T., {De Ridder}, J., {et~al.} 2022{\natexlab{b}},
  \aap, 662, A82, \dodoi{10.1051/0004-6361/202141926}

\bibitem[{Guzik(2021)}]{Guzik2021Highlights}
Guzik, J.~A. 2021, Frontiers in Astronomy and Space Sciences, 8,
  \dodoi{10.3389/fspas.2021.653558}

\bibitem[{{Guzik} {et~al.}(2000){Guzik}, {Kaye}, {Bradley}, {Cox}, \&
  {Neuforge}}]{Guzik2000}
{Guzik}, J.~A., {Kaye}, A.~B., {Bradley}, P.~A., {Cox}, A.~N., \& {Neuforge},
  C. 2000, \apjl, 542, L57, \dodoi{10.1086/312908}

\bibitem[{Han \& Brandt(2023)}]{Han2023TESSGaia}
Han, T., \& Brandt, T.~D. 2023, The Astronomical Journal, 165, 71,
  \dodoi{10.3847/1538-3881/acaaa7}

\bibitem[{Handler {et~al.}(2019)Handler, Pigulski, {Daszy{\'n}ska-Daszkiewicz},
  Irrgang, Kilkenny, Guo, Przybilla, Kahraman~Ali{\c c}avu{\c s}, Kallinger,
  {Pascual-Granado}, Niemczura, R{\'o}{\.z}a{\'n}ski, Chowdhury, Buzasi,
  Mirouh, Bowman, Johnston, Pedersen, {Sim{\'o}n-D{\'i}az}, Moravveji, Gazeas,
  De~Cat, Vanderspek, \& Ricker}]{Handler2019Asteroseismology}
Handler, G., Pigulski, A., {Daszy{\'n}ska-Daszkiewicz}, J., {et~al.} 2019, The
  Astrophysical Journal, 873, L4, \dodoi{10.3847/2041-8213/ab095f}

\bibitem[{{Hekker} \& {Christensen-Dalsgaard}(2017)}]{HekkerJCD2017}
{Hekker}, S., \& {Christensen-Dalsgaard}, J. 2017, \aapr, 25, 1,
  \dodoi{10.1007/s00159-017-0101-x}

\bibitem[{{Hermes} {et~al.}(2017){Hermes}, {G{\"a}nsicke}, {Kawaler}, {Greiss},
  {Tremblay}, {Gentile Fusillo}, {Raddi}, {Fanale}, {Bell}, {Dennihy}, {Fuchs},
  {Dunlap}, {Clemens}, {Montgomery}, {Winget}, {Chote}, {Marsh}, \&
  {Redfield}}]{Hermes2017}
{Hermes}, J.~J., {G{\"a}nsicke}, B.~T., {Kawaler}, S.~D., {et~al.} 2017, \apjs,
  232, 23, \dodoi{10.3847/1538-4365/aa8bb5}

\bibitem[{Hey {et~al.}(2021)Hey, Montet, Pope, Murphy, \&
  Bedding}]{Hey2021Searcha}
Hey, D.~R., Montet, B.~T., Pope, B. J.~S., Murphy, S.~J., \& Bedding, T.~R.
  2021, arXiv:2108.03785 [astro-ph].
\newblock \doeprint{2108.03785}

\bibitem[{Hey {et~al.}(2019)Hey, Holdsworth, Bedding, Murphy, Cunha, Kurtz,
  Huber, Fulton, \& Howard}]{Hey2019Six}
Hey, D.~R., Holdsworth, D.~L., Bedding, T.~R., {et~al.} 2019, Monthly Notices
  of the Royal Astronomical Society, 488, 18, \dodoi{10.1093/mnras/stz1633}

\bibitem[{{Heynderickx} {et~al.}(1994){Heynderickx}, {Waelkens}, \&
  {Smeyers}}]{Heynderickx1994}
{Heynderickx}, D., {Waelkens}, C., \& {Smeyers}, P. 1994, \aaps, 105, 447

\bibitem[{{Holl} {et~al.}(2018){Holl}, {Audard}, {Nienartowicz}, {Jevardat de
  Fombelle}, {Marchal}, {Mowlavi}, {Clementini}, {De Ridder}, {Evans}, {Guy},
  {Lanzafame}, {Lebzelter}, {Rimoldini}, {Roelens}, {Zucker}, {Distefano},
  {Garofalo}, {Lecoeur-Ta{\"\i}bi}, {Lopez}, {Molinaro}, {Muraveva}, {Panahi},
  {Regibo}, {Ripepi}, {Sarro}, {Aerts}, {Anderson}, {Charnas}, {Barblan},
  {Blanco-Cuaresma}, {Busso}, {Cuypers}, {De Angeli}, {Glass}, {Grenon},
  {Juh{\'a}sz}, {Kochoska}, {Koubsky}, {Lanza}, {Leccia}, {Lorenz}, {Marconi},
  {Marschalk{\'o}}, {Mazeh}, {Messina}, {Mignard}, {Moitinho}, {Moln{\'a}r},
  {Morgenthaler}, {Musella}, {Ordenovic}, {Ord{\'o}{\~n}ez}, {Pagano},
  {Palaversa}, {Pawlak}, {Plachy}, {Pr{\v{s}}a}, {Riello}, {S{\"u}veges},
  {Szabados}, {Szegedi-Elek}, {Votruba}, \& {Eyer}}]{Holl2018}
{Holl}, B., {Audard}, M., {Nienartowicz}, K., {et~al.} 2018, \aap, 618, A30,
  \dodoi{10.1051/0004-6361/201832892}

\bibitem[{Huang {et~al.}(2020)Huang, Vanderburg, P{\'a}l, Sha, Yu, Fong,
  Fausnaugh, Shporer, Guerrero, Vanderspek, \& Ricker}]{Huang2020Photometry}
Huang, C.~X., Vanderburg, A., P{\'a}l, A., {et~al.} 2020, Photometry of 10
  {{Million Stars}} from the {{First Two Years}} of {{TESS Full Frame Images}},
  \dodoi{10.17909/t9-r086-e880}

\bibitem[{{IJspeert} {et~al.}(2021){IJspeert}, {Tkachenko}, {Johnston},
  {Garcia}, {De Ridder}, {Van Reeth}, \& {Aerts}}]{IJspeert2021}
{IJspeert}, L.~W., {Tkachenko}, A., {Johnston}, C., {et~al.} 2021, \aap, 652,
  A120, \dodoi{10.1051/0004-6361/202141489}

\bibitem[{{IJspeert} {et~al.}(2024){IJspeert}, {Tkachenko}, {Johnston},
  {Pr{\v{s}}a}, {Wells}, \& {Aerts}}]{IJspeert2024}
---. 2024, \aap, in press, arXiv:2402.06084, \dodoi{10.48550/arXiv.2402.06084}

\bibitem[{{Johnston}(2021)}]{Johnston2021}
{Johnston}, C. 2021, \aap, 655, A29, \dodoi{10.1051/0004-6361/202141080}

\bibitem[{{Kirk} {et~al.}(2016){Kirk}, {Conroy}, {Pr{\v{s}}a}, {Abdul-Masih},
  {Kochoska}, {Matijevi{\v{c}}}, {Hambleton}, {Barclay}, {Bloemen}, {Boyajian},
  {Doyle}, {Fulton}, {Hoekstra}, {Jek}, {Kane}, {Kostov}, {Latham}, {Mazeh},
  {Orosz}, {Pepper}, {Quarles}, {Ragozzine}, {Shporer}, {Southworth},
  {Stassun}, {Thompson}, {Welsh}, {Agol}, {Derekas}, {Devor}, {Fischer},
  {Green}, {Gropp}, {Jacobs}, {Johnston}, {LaCourse}, {Saetre}, {Schwengeler},
  {Toczyski}, {Werner}, {Garrett}, {Gore}, {Martinez}, {Spitzer}, {Stevick},
  {Thomadis}, {Vrijmoet}, {Yenawine}, {Batalha}, \& {Borucki}}]{Kirk2016}
{Kirk}, B., {Conroy}, K., {Pr{\v{s}}a}, A., {et~al.} 2016, \aj, 151, 68,
  \dodoi{10.3847/0004-6256/151/3/68}

\bibitem[{{Kurtz}(2022)}]{Kurtz2022}
{Kurtz}, D.~W. 2022, \araa, 60, 31, \dodoi{10.1146/annurev-astro-052920-094232}

\bibitem[{Kurtz {et~al.}(2015)Kurtz, Shibahashi, Murphy, Bedding, \&
  Bowman}]{Kurtz2015unifying}
Kurtz, D.~W., Shibahashi, H., Murphy, S.~J., Bedding, T.~R., \& Bowman, D.~M.
  2015, Monthly Notices of the Royal Astronomical Society, 450, 3015,
  \dodoi{10.1093/mnras/stv868}

\bibitem[{{Li} {et~al.}(2019){Li}, {Bedding}, {Murphy}, {Van Reeth}, {Antoci},
  \& {Ouazzani}}]{GangLi2019}
{Li}, G., {Bedding}, T.~R., {Murphy}, S.~J., {et~al.} 2019, \mnras, 482, 1757,
  \dodoi{10.1093/mnras/sty2743}

\bibitem[{{Li} {et~al.}(2020){Li}, {Van Reeth}, {Bedding}, {Murphy}, {Antoci},
  {Ouazzani}, \& {Barbara}}]{GangLi2020}
{Li}, G., {Van Reeth}, T., {Bedding}, T.~R., {et~al.} 2020, \mnras, 491, 3586,
  \dodoi{10.1093/mnras/stz2906}

\bibitem[{Li {et~al.}(2020)Li, Van~Reeth, Bedding, Murphy, Antoci, Ouazzani, \&
  Barbara}]{Li2020Gravitymode}
Li, G., Van~Reeth, T., Bedding, T.~R., {et~al.} 2020, Monthly Notices of the
  Royal Astronomical Society, 491, 3586, \dodoi{10.1093/mnras/stz2906}

\bibitem[{{Li} {et~al.}(2024){Li}, {Aerts}, {Bedding}, {Fritzewski}, {Murphy},
  {Van Reeth}, {Montet}, {Jian}, {Mombarg}, {Gossage}, \&
  {Sreenivas}}]{GangLi2024}
{Li}, G., {Aerts}, C., {Bedding}, T.~R., {et~al.} 2024, \aap, submitted,
  arXiv:2311.16991, \dodoi{10.48550/arXiv.2311.16991}

\bibitem[{Lubba {et~al.}(2019)Lubba, Sethi, Knaute, Schultz, Fulcher, \&
  Jones}]{Lubba2019Catch22}
Lubba, C.~H., Sethi, S.~S., Knaute, P., {et~al.} 2019, Data Mining and
  Knowledge Discovery, 33, 1821, \dodoi{10.1007/s10618-019-00647-x}

\bibitem[{{Mombarg} {et~al.}(2024){Mombarg}, {Aerts}, \&
  {Molenberghs}}]{Mombarg2024}
{Mombarg}, J.~S.~G., {Aerts}, C., \& {Molenberghs}, G. 2024, \aap, 685, A21,
  \dodoi{10.1051/0004-6361/202449213}

\bibitem[{{Mombarg} {et~al.}(2021){Mombarg}, {Van Reeth}, \&
  {Aerts}}]{Mombarg2021}
{Mombarg}, J.~S.~G., {Van Reeth}, T., \& {Aerts}, C. 2021, \aap, 650, A58,
  \dodoi{10.1051/0004-6361/202039543}

\bibitem[{{Murphy} {et~al.}(2022){Murphy}, {Bedding}, {White}, {Li}, {Hey},
  {Reese}, \& {Joyce}}]{Murphy2022}
{Murphy}, S.~J., {Bedding}, T.~R., {White}, T.~R., {et~al.} 2022, \mnras, 511,
  5718, \dodoi{10.1093/mnras/stac240}

\bibitem[{Murphy {et~al.}(2019)Murphy, Hey, Van~Reeth, \&
  Bedding}]{Murphy2019Gaiaderived}
Murphy, S.~J., Hey, D., Van~Reeth, T., \& Bedding, T.~R. 2019, Monthly Notices
  of the Royal Astronomical Society, 485, 2380, \dodoi{10.1093/mnras/stz590}

\bibitem[{Murphy {et~al.}(2013)Murphy, Shibahashi, \&
  Kurtz}]{Murphy2013SuperNyquist}
Murphy, S.~J., Shibahashi, H., \& Kurtz, D.~W. 2013, Monthly Notices of the
  Royal Astronomical Society, 430, 2986, \dodoi{10.1093/mnras/stt105}

\bibitem[{Netzel {et~al.}(2022)Netzel, Pietrukowicz, Soszy{\'n}ski, \&
  Wrona}]{Netzel2022Frequency}
Netzel, H., Pietrukowicz, P., Soszy{\'n}ski, I., \& Wrona, M. 2022, Monthly
  Notices of the Royal Astronomical Society, 510, 1748,
  \dodoi{10.1093/mnras/stab3555}

\bibitem[{{Pamyatnykh}(1999)}]{Pamyatnykh1999}
{Pamyatnykh}, A.~A. 1999, \actaa, 49, 119

\bibitem[{{Panda} {et~al.}(2024){Panda}, {Dhanpal}, {Murphy}, {Hanasoge}, \&
  {Bedding}}]{Panda2024}
{Panda}, S.~K., {Dhanpal}, S., {Murphy}, S.~J., {Hanasoge}, S., \& {Bedding},
  T.~R. 2024, \apj, 960, 94, \dodoi{10.3847/1538-4357/ad0a97}

\bibitem[{{Pedersen}(2022{\natexlab{a}})}]{Pedersen2022b}
{Pedersen}, M.~G. 2022{\natexlab{a}}, \apj, 940, 49,
  \dodoi{10.3847/1538-4357/ac947f}

\bibitem[{{Pedersen}(2022{\natexlab{b}})}]{Pedersen2022}
---. 2022{\natexlab{b}}, \apj, 930, 94, \dodoi{10.3847/1538-4357/ac5b05}

\bibitem[{{Pedersen} {et~al.}(2017){Pedersen}, {Antoci}, {Korhonen}, {White},
  {Jessen-Hansen}, {Lehtinen}, {Nikbakhsh}, \& {Viuho}}]{Pedersen2017}
{Pedersen}, M.~G., {Antoci}, V., {Korhonen}, H., {et~al.} 2017, \mnras, 466,
  3060, \dodoi{10.1093/mnras/stw3226}

\bibitem[{{Pedersen} {et~al.}(2019){Pedersen}, {Chowdhury}, {Johnston},
  {Bowman}, {Aerts}, {Handler}, {De Cat}, {Neiner}, {David-Uraz}, {Buzasi},
  {Tkachenko}, {Sim{\'o}n-D{\'\i}az}, {Moravveji}, {Sikora}, {Mirouh},
  {Lovekin}, {Cantiello}, {Daszy{\'n}ska-Daszkiewicz}, {Pigulski},
  {Vanderspek}, \& {Ricker}}]{Pedersen2019}
{Pedersen}, M.~G., {Chowdhury}, S., {Johnston}, C., {et~al.} 2019, \apjl, 872,
  L9, \dodoi{10.3847/2041-8213/ab01e1}

\bibitem[{{Pedersen} {et~al.}(2021){Pedersen}, {Aerts}, {P{\'a}pics},
  {Michielsen}, {Gebruers}, {Rogers}, {Molenberghs}, {Burssens}, {Garcia}, \&
  {Bowman}}]{Pedersen2021}
{Pedersen}, M.~G., {Aerts}, C., {P{\'a}pics}, P.~I., {et~al.} 2021, Nature
  Astronomy, 5, 715, \dodoi{10.1038/s41550-021-01351-x}

\bibitem[{Pietrukowicz {et~al.}(2020)Pietrukowicz, Soszynski, Netzel, Wrona,
  Udalski, Szymanski, Poleski, Kozlowski, Skowron, Ulaczyk, Skowron, Mroz,
  Rybicki, Iwanek, \& Gromadzki}]{Pietrukowicz202010}
Pietrukowicz, P., Soszynski, I., Netzel, H., {et~al.} 2020, Acta Astronomica,
  70, 241, \dodoi{10.32023/0001-5237/70.4.1}

\bibitem[{{Rauer et al.}(2024)}]{Rauer2024}
{Rauer et al.} 2024, Experimental Astronomy, submitted

\bibitem[{{Rehm} {et~al.}(2024){Rehm}, {Mombarg}, {Aerts}, {Michielsen},
  {Burssens}, \& {Townsend}}]{Rehm2024}
{Rehm}, R., {Mombarg}, J.~S.~G., {Aerts}, C., {et~al.} 2024, \aap, in press,
  arXiv:2405.08864, \dodoi{10.48550/arXiv.2405.08864}

\bibitem[{{Ricker} {et~al.}(2015){Ricker}, {Winn}, {Vanderspek}, {Latham},
  {Bakos}, {Bean}, {Berta-Thompson}, {Brown}, {Buchhave}, {Butler}, {Butler},
  {Chaplin}, {Charbonneau}, {Christensen-Dalsgaard}, {Clampin}, {Deming},
  {Doty}, {De Lee}, {Dressing}, {Dunham}, {Endl}, {Fressin}, {Ge}, {Henning},
  {Holman}, {Howard}, {Ida}, {Jenkins}, {Jernigan}, {Johnson}, {Kaltenegger},
  {Kawai}, {Kjeldsen}, {Laughlin}, {Levine}, {Lin}, {Lissauer}, {MacQueen},
  {Marcy}, {McCullough}, {Morton}, {Narita}, {Paegert}, {Palle}, {Pepe},
  {Pepper}, {Quirrenbach}, {Rinehart}, {Sasselov}, {Sato}, {Seager},
  {Sozzetti}, {Stassun}, {Sullivan}, {Szentgyorgyi}, {Torres}, {Udry}, \&
  {Villasenor}}]{Ricker2015}
{Ricker}, G.~R., {Winn}, J.~N., {Vanderspek}, R., {et~al.} 2015, Journal of
  Astronomical Telescopes, Instruments, and Systems, 1, 014003,
  \dodoi{10.1117/1.JATIS.1.1.014003}

\bibitem[{{Rimoldini} {et~al.}(2023){Rimoldini}, {Holl}, {Gavras}, {Audard},
  {De Ridder}, {Mowlavi}, {Nienartowicz}, {Jevardat de Fombelle},
  {Lecoeur-Ta{\"\i}bi}, {Karbevska}, {Evans}, {{\'A}brah{\'a}m}, {Carnerero},
  {Clementini}, {Distefano}, {Garofalo}, {Garc{\'\i}a-Lario}, {Gomel},
  {Klioner}, {Kruszy{\'n}ska}, {Lanzafame}, {Lebzelter}, {Marton}, {Mazeh},
  {Molinaro}, {Panahi}, {Raiteri}, {Ripepi}, {Szabados}, {Teyssier},
  {Trabucchi}, {Wyrzykowski}, {Zucker}, \& {Eyer}}]{Rimoldi2023}
{Rimoldini}, L., {Holl}, B., {Gavras}, P., {et~al.} 2023, \aap, 674, A14,
  \dodoi{10.1051/0004-6361/202245591}

\bibitem[{Shi {et~al.}(2023)Shi, Qian, Zhu, \& Li}]{Shi2023Catalog}
Shi, X.-d., Qian, S.-b., Zhu, L.-y., \& Li, L.-j. 2023, The Astrophysical
  Journal Supplement Series, 268, 16, \dodoi{10.3847/1538-4365/ace88c}

\bibitem[{{Shultz} {et~al.}(2019){Shultz}, {Wade}, {Rivinius}, {Alecian},
  {Neiner}, {Petit}, {Owocki}, {ud-Doula}, {Kochukhov}, {Bohlender},
  {Keszthelyi}, {MiMeS Collaboration}, \& {BinaMIcS
  Collaboration}}]{Shultz2019}
{Shultz}, M.~E., {Wade}, G.~A., {Rivinius}, T., {et~al.} 2019, \mnras, 490,
  274, \dodoi{10.1093/mnras/stz2551}

\bibitem[{{Sikora} {et~al.}(2019{\natexlab{a}}){Sikora}, {Wade}, {Power}, \&
  {Neiner}}]{Sikora2019a}
{Sikora}, J., {Wade}, G.~A., {Power}, J., \& {Neiner}, C. 2019{\natexlab{a}},
  \mnras, 483, 3127, \dodoi{10.1093/mnras/sty2895}

\bibitem[{{Sikora} {et~al.}(2019{\natexlab{b}}){Sikora}, {David-Uraz},
  {Chowdhury}, {Bowman}, {Wade}, {Khalack}, {Kobzar}, {Kochukhov}, {Neiner}, \&
  {Paunzen}}]{Sikora2019b}
{Sikora}, J., {David-Uraz}, A., {Chowdhury}, S., {et~al.} 2019{\natexlab{b}},
  \mnras, 487, 4695, \dodoi{10.1093/mnras/stz1581}

\bibitem[{Skarka {et~al.}(2022)Skarka, {\v Z}{\'a}k, Fedurco, Paunzen, Henzl,
  Ma{\v s}ek, Karjalainen, Arias, S{\'o}dor, Auer, Kab{\'a}th, Karjalainen,
  Li{\v s}ka, \& {\v S}tegner}]{Skarka2022Periodic}
Skarka, M., {\v Z}{\'a}k, J., Fedurco, M., {et~al.} 2022, Astronomy \&
  Astrophysics, 666, A142, \dodoi{10.1051/0004-6361/202244037}

\bibitem[{{Stankov} \& {Handler}(2005)}]{Stankov2005}
{Stankov}, A., \& {Handler}, G. 2005, \apjs, 158, 193, \dodoi{10.1086/429408}

\bibitem[{Stassun {et~al.}(2018)Stassun, Oelkers, Pepper, Paegert, Lee, Torres,
  Latham, Charpinet, Dressing, Huber, Kane, L{\'e}pine, Mann, Muirhead,
  {Rojas-Ayala}, Silvotti, Fleming, Levine, \& Plavchan}]{Stassun2018TESS}
Stassun, K.~G., Oelkers, R.~J., Pepper, J., {et~al.} 2018, The Astronomical
  Journal, 156, 102, \dodoi{10.3847/1538-3881/aad050}

\bibitem[{Steen {et~al.}(2024)Steen, Hermes, Guidry, Paiva, Farihi, Heintz,
  Ewing, \& Berry}]{Steen2024Measuring}
Steen, M., Hermes, J.~J., Guidry, J.~A., {et~al.} 2024, Measuring {{White Dwarf
  Variability}} from {{Sparsely Sampled Gaia DR3 Multi-Epoch Photometry}},
  \dodoi{10.48550/arXiv.2404.02201}

\bibitem[{{Szewczuk} \& {Daszy{\'n}ska-Daszkiewicz}(2017)}]{Szewczuk2017}
{Szewczuk}, W., \& {Daszy{\'n}ska-Daszkiewicz}, J. 2017, \mnras, 469, 13,
  \dodoi{10.1093/mnras/stx738}

\bibitem[{{Townsend}(2003)}]{Townsend2003}
{Townsend}, R.~H.~D. 2003, \mnras, 340, 1020,
  \dodoi{10.1046/j.1365-8711.2003.06379.x}

\bibitem[{{Townsend}(2005)}]{Townsend2005}
---. 2005, \mnras, 364, 573, \dodoi{10.1111/j.1365-2966.2005.09585.x}

\bibitem[{{Uytterhoeven} {et~al.}(2011){Uytterhoeven}, {Moya},
  {Grigahc{\`e}ne}, {Guzik}, {Guti{\'e}rrez-Soto}, {Smalley}, {Handler},
  {Balona}, {Niemczura}, {Fox Machado}, {Benatti}, {Chapellier}, {Tkachenko},
  {Szab{\'o}}, {Su{\'a}rez}, {Ripepi}, {Pascual}, {Mathias},
  {Mart{\'\i}n-Ru{\'\i}z}, {Lehmann}, {Jackiewicz}, {Hekker}, {Gruberbauer},
  {Garc{\'\i}a}, {Dumusque}, {D{\'\i}az-Fraile}, {Bradley}, {Antoci}, {Roth},
  {Leroy}, {Murphy}, {De Cat}, {Cuypers}, {Kjeldsen}, {Christensen-Dalsgaard},
  {Breger}, {Pigulski}, {Kiss}, {Still}, {Thompson}, \& {van
  Cleve}}]{Uytterhoeven2011}
{Uytterhoeven}, K., {Moya}, A., {Grigahc{\`e}ne}, A., {et~al.} 2011, \aap, 534,
  A125, \dodoi{10.1051/0004-6361/201117368}

\bibitem[{{Van Beeck} {et~al.}(2021){Van Beeck}, {Bowman}, {Pedersen}, {Van
  Reeth}, {Van Hoolst}, \& {Aerts}}]{VanBeeck2021}
{Van Beeck}, J., {Bowman}, D.~M., {Pedersen}, M.~G., {et~al.} 2021, \aap, 655,
  A59, \dodoi{10.1051/0004-6361/202141572}

\bibitem[{{Van Reeth} {et~al.}(2016){Van Reeth}, {Tkachenko}, \&
  {Aerts}}]{VanReeth2016}
{Van Reeth}, T., {Tkachenko}, A., \& {Aerts}, C. 2016, \aap, 593, A120,
  \dodoi{10.1051/0004-6361/201628616}

\bibitem[{{Van Reeth} {et~al.}(2015){Van Reeth}, {Tkachenko}, {Aerts},
  {P{\'a}pics}, {Triana}, {Zwintz}, {Degroote}, {Debosscher}, {Bloemen},
  {Schmid}, {De Smedt}, {Fremat}, {Fuentes}, {Homan}, {Hrudkova},
  {Karjalainen}, {Lombaert}, {Nemeth}, {{\O}stensen}, {Van De Steene}, {Vos},
  {Raskin}, \& {Van Winckel}}]{VanReeth2015}
{Van Reeth}, T., {Tkachenko}, A., {Aerts}, C., {et~al.} 2015, \apjs, 218, 27,
  \dodoi{10.1088/0067-0049/218/2/27}

\bibitem[{{Van Reeth} {et~al.}(2018){Van Reeth}, {Mombarg}, {Mathis},
  {Tkachenko}, {Fuller}, {Bowman}, {Buysschaert}, {Johnston}, {Garc{\'\i}a
  Hern{\'a}ndez}, {Goldstein}, {Townsend}, \& {Aerts}}]{VanReeth2018}
{Van Reeth}, T., {Mombarg}, J.~S.~G., {Mathis}, S., {et~al.} 2018, \aap, 618,
  A24, \dodoi{10.1051/0004-6361/201832718}

\bibitem[{{Weiler}(2018)}]{Weiler2018}
{Weiler}, M. 2018, \aap, 617, A138, \dodoi{10.1051/0004-6361/201833462}

\bibitem[{{White} {et~al.}(2017){White}, {Pope}, {Antoci}, {P{\'a}pics},
  {Aerts}, {Gies}, {Gordon}, {Huber}, {Schaefer}, {Aigrain}, {Albrecht},
  {Barclay}, {Barentsen}, {Beck}, {Bedding}, {Fredslund Andersen}, {Grundahl},
  {Howell}, {Ireland}, {Murphy}, {Nielsen}, {Silva Aguirre}, \&
  {Tuthill}}]{White2017}
{White}, T.~R., {Pope}, B.~J.~S., {Antoci}, V., {et~al.} 2017, \mnras, 471,
  2882, \dodoi{10.1093/mnras/stx1050}

\end{thebibliography}

\begin{appendix} 
    \section{Example classified light curves}
    \nopagebreak
    \begin{figure}[b!]
        \begin{center}
            \includegraphics[]{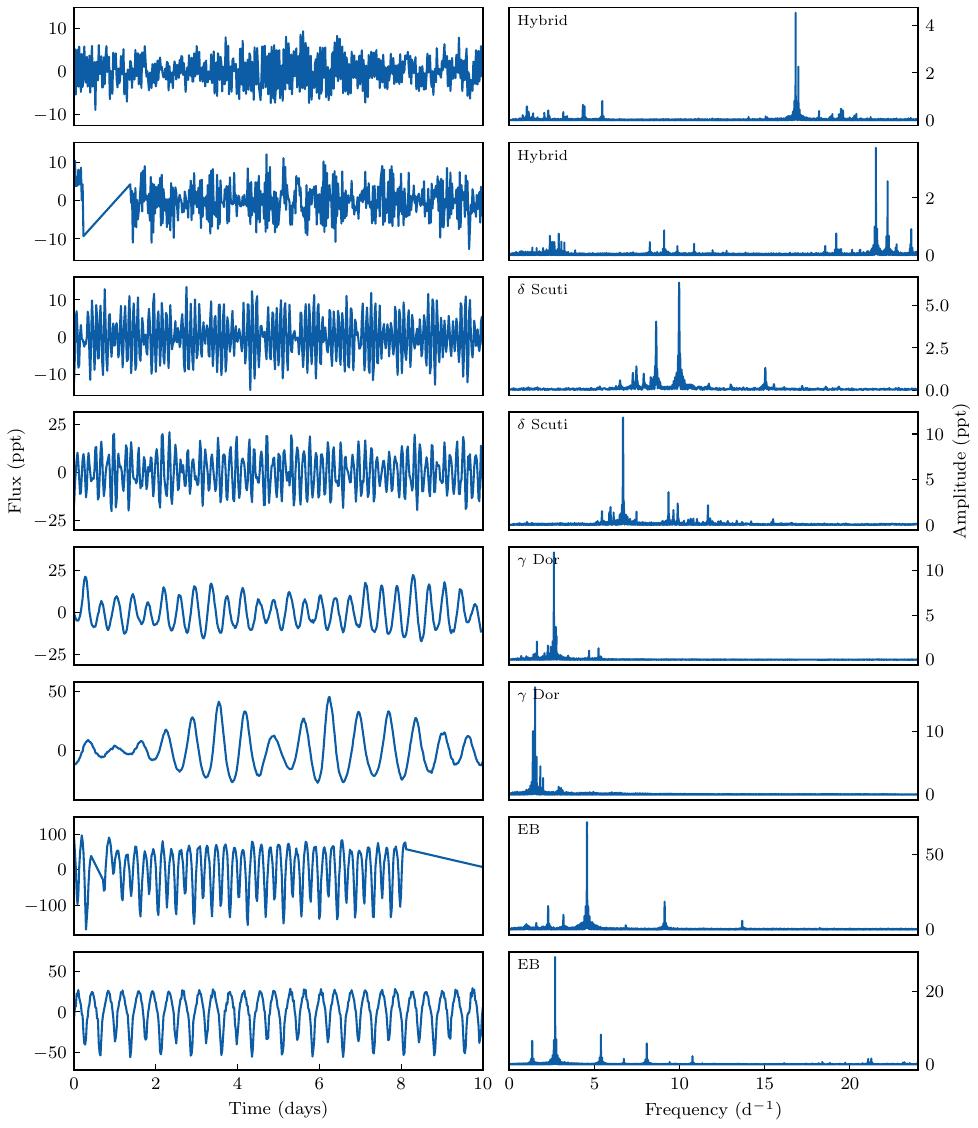}
        \end{center}
        \caption{\label{example_class} A series of prototypical classified light curves from Sec.~4.2, where the classification probability is greater than 0.8. Note that we have stitched multiple TESS sectors to increase the frequency resolution.}
    \end{figure}

\end{appendix}

\end{document}